\documentclass[a4paper,fleqn]{cas-dc}

\usepackage[numbers]{natbib}
\usepackage{amsmath,amsthm,amssymb}
\usepackage{booktabs}
\usepackage{hyperref}
\usepackage{cleveref}
\usepackage{graphicx,subcaption}
\usepackage{comment}
\usepackage{lipsum}
\usepackage{soul}

\def\tsc#1{\csdef{#1}{\textsc{\lowercase{#1}}\xspace}}
\tsc{WGM}
\tsc{QE}
\newcommand{\xmark}{\text{\sffamily X}}%
\begin{document}
\let\WriteBookmarks\relax
\def\floatpagepagefraction{1}
\def\textpagefraction{.001}

\shorttitle{Vector self-interaction  and Neutron Star properties}    

\shortauthors{B. K. Pradhan,D. Chatterjee, R. Gandhi, J.  Schaffner-Bielich}  

\title [mode = title]{Role of vector self-interaction in Neutron Star properties}  

%\tnotemark[<tnote number>] 

%\tnotetext[<tnote number>]{<tnote text>} 


\author[1]{Bikram Keshari Pradhan}[orcid=0000-0002-2526-1421]
\cormark[1]
\ead{bikramp@iucaa.in}

%\ead[url]{<URL>}

\credit{Code development, Methodology, Software, Preparing Manuscript}

\affiliation[1]{organization={Inter University Center for Astronomy and Astrophysics},
            city={Pune},
%          citysep={}, % Uncomment if no comma needed between city and postcode
            postcode={411007}, 
           % state={Maharastra},
            country={India}}

\author[1]{Debarati Chatterjee}[orcid=0000-0002-0995-2329]

% Footnote of the second author
%\fnmark[]

% Email id of the second author
\ead{debarati@iucaa.in}
\credit{Conceptualization of this study, code checking, Methodology, Preparing Manuscript}
\author[2]{Radhika Gandhi}
\affiliation[2]{organization={Indian Institute of Technology Gandhinagar},
            addressline={Palaj}, 
            city={Gandhinagar},
%          citysep={}, % Uncomment if no comma needed between city and postcode
            postcode={382055}, 
 %           state={},
            country={India}}

% Footnote of the second author
%\fnmark[]

% Email id of the second author

\credit{Code Checking,Preparing Manuscript}

\author[3]{J\"urgen Schaffner-Bielich}

\affiliation[3]{organization={Institut f\"ur Theoretische Physik,
Goethe Universit\"at},
            addressline={Max von Laue Str. 1}, 
           city={Frankfurt am Main},
%          citysep={}, % Uncomment if no comma needed between city and postcode
            postcode={60438 }, 
 %           state={},
            country={ Germany}}
 \credit{Conceptualization of this study, Methodology, Preparing the Manuscript}
\cortext[1]{Bikram Keshari Pradhan}

%\fntext[]{}


\begin{abstract}
Previous studies have claimed that there exist correlations among certain nuclear saturation parameters and neutron star observables, such as the slope of the symmetry energy and the radius of a $1.4M_{\odot}$ neutron star. However, it is not clear whether such correlations are physical or spurious, as they are not observed universally for all equation of state models. In this work, we probe the role of vector self-interaction within the framework of the Relativistic Mean Field model and its role in governing the observable stellar properties and their correlations with nuclear parameters. We confirm that the effect of this term is not only to control the high density properties of the equation of state but also in governing such correlations. We also impose a limit on the maximum strength of the vector self-interaction using recent astrophysical data.
\end{abstract}

%%includegraphics{}

%\item 
%\item 
%\item 

% Keywords
\begin{keywords}
 \sep Neutron Star \sep Equation of State \sep RMF Model \sep Meson interaction 
\end{keywords}

\maketitle

\section{Introduction}
\label{sec:intro}

%\dc{Neutron stars: what they are, why we study them, dense matter : 1 para} \\
Neutron stars (NSs) are compact remnants of stellar evolution, that allow us to study extreme matter physics that is beyond the reach of terrestrial experiments. While nuclear experiments give us information about the nuclear interaction close to nuclear saturation density ($\rho_0 \sim 10^{17}$ kg/m$^3$), densities in the core of a neutron star can reach much higher values. Heavy-ion collision experiments in particle accelerators can reach densities up to several times $\rho_0$, but both heavy-ion and nuclear experiments probe approximately symmetric nuclear matter (equal number of neutrons and protons) while neutron stars are highly isospin-asymmetric. This introduces uncertainties due to the extrapolation of our current knowledge of nuclear interactions to unknown territories of high densities and asymmetries~\cite{Lattimer_2001,Vidana2020} .
\\

%\dc{NS observables - multi-messenger astronomy (EM+GW) : 1 para}\\
A recent breakthrough in this field has been achieved by multi-messenger astrophysical observations of NSs. While these compact objects have been observed for many decades at multiple frequencies across the electromagnetic spectrum using ground- and space-based facilities, a new method to directly probe its interior has emerged with the recent direct detection of gravitational waves by the LIGO-Virgo Collaboration ~\cite{Abbott2017,Abbott_2017_apjl}. In particular, the detection of GWs from the binary NS merger GW170817 and its counterparts in electromagnetic observations have led to the verification of many theoretical conjectures related to these systems as well as major implications for the study of the equation of state (EoS) of dense matter~\cite{Abbott2017,Abbott2019}. Using this wealth of astrophysical data, it is possible to derive a number of global stellar quantities, such as their mass, radius, or tidal deformation, all of which depend on the NS EoS.
\\

%\dc{Nuclear observables, EoS models review, RMF model : 1 para}\\
The EoS relates the underlying nuclear interactions to the global properties of the NS.  There are two main approaches to describing the nuclear EoS~\cite{OertelRMP}; the first is microscopic or {\it ab-initio} while the other includes  phenomenological (effective theories where parameters are fitted to reproduce the saturation properties of nuclear matter and/or the properties of finite nuclei)  interactions of Skyrme
\cite{Skyrme1956,Skyrme1956b,CHABANAT1997,Sagawa2007,STONE2007587,Vautherin1972}, Gogny type ~\cite{Decharg1980,BergerNPA1989,Sellahewa2014,Chappert2015, Chen2012} and those of the relativistic
approach~\cite{MACHLEIDT19871,Haidenbauer,Shen2010,Cescato1998,Nik2002,book1,Serot1997}. In this work, we employ the Relativistic Mean Field (RMF) model, which is a relativistic effective field theoretical model successfully applied to a wide range of nuclei and nuclear matter~\cite{Chen2014,Hornick,Pradhan2021,Ghosh2022,Ghosh2022b,Thakur2022}. 
\\

%\dc{Aim of this work: why study the vector self-interaction term - 1 para}\\
Previous studies in the literature have shown that there may exist correlations among certain nuclear saturation parameters and neutron star observables, such as the slope of the symmetry energy and the radius of a ${1.4 M_{\odot}}$ neutron star ~\cite{Fattoyev,Zhang2019,Zhang_2020,Gueven2020,Fattoyev2018,Tsang2020,Hu2020,Malik2020}. However, such correlations are not observed for some other EoS models \cite{Hornick,Ghosh2022,Ghosh2022b,Malik2018,biswas2021prex,Reed2021,Pradhan2022}, it is not clear whether these are physical or spurious.  In this work, we particularly probe the vector self-interaction within the framework of the RMF model and its role in governing the observable stellar properties (mass, radius, tidal deformability) and their correlations with nuclear parameters. 
\\

%\dc{Structure of the paper}\\
The outline of the paper is as follows: in ~\Cref{sec:formalism}, we describe the methodology of the microscopic EoS and the global structure of the NS. In ~\Cref{sec:results}, we provide the results of our investigation. Finally, in ~\Cref{sec:discussions}, we discuss the implications of the results of this study. In this work, we adopt the natural unit system with $\hbar=c=G=1$.
\\
\section{Formalism}
\label{sec:formalism}
As mentioned in ~\Cref{sec:intro}, different theoretical models have been developed in order to describe the behavior of nuclear matter at high densities. In Relativistic Mean Field (RMF) models, the strong force between nucleons is mediated by the exchange of mesons. It is a phenomenological model in which each particle  feels the potential of other particles, and this framework of mean-field approximation uses the expectation values removing all quantum fluctuations of the meson fields. In this study, for the EoS we consider RMF models with certain nonlinear meson self-couplings.

Consequently, the RMF model parameter set is obtained by simultaneously fitting the isoscalar couplings to saturation nuclear properties and nucleon effective mass while allowing for variations of the isovector parameters  so as to reproduce the symmetry energy and its slope within reasonable theoretical and experimental limits.  For this investigation, we  follow Hornick et al.~\cite{Hornick}, which allows for variation of the parameter space within the current uncertainties in nuclear empirical observables. The resultant parameter set, including the ranges of saturation number density $n_{0}$, 
energy per particle $E_{sat}$, the incompressibility coefficient $K_{sat}$,
the effective mass of nucleon $m^*/m_N$,
symmetry energy $J_{sym}$ 
 and slope of symmetry energy $L$  is represented in ~\Cref{tab:HTZCS_parameter_set}.
 
 \subsection{Microscopic EoS model}
\label{sec: eos}

%\dc{RMF model details, equations}
Equation of state plays a vital role in relating nuclear matter and properties of the NS. For this investigation, we adopt the relativistic mean field  (RMF) model   to describe the $\beta-$equilibrated and charge-neutral hadronic matter. The NS core is assumed to be composed of nucleons (neutrons and protons) and leptons (electrons $e^-$ and muons $\mu^-$). In RMF theory, the Lagrangian density describes the interaction between baryons through the exchange of mesons: the scalar-isoscalar ($\sigma$), vector-isoscalar ($\omega$), vector-isovector ($\rho$) mesons as given in Eq.~\eqref{eqn:lagr}. In addition, the Lagrangian density takes into account the scalar and vector self-interactions, while also including contributions from possible mixed interactions between the mesons up to quartic order. 
\begin{eqnarray}
     \mathcal{L} &=&\sum_N  \bar{\psi}_{N}  (i\gamma^{\mu}\partial_{\mu}-m +g_{\sigma}\sigma-g_{\omega B}\gamma_{\mu}\omega^{\mu}-\frac{g_{\rho}}{2}\gamma_{\mu} \vec{\tau}\vec{\rho}^{\mu})\psi_{N} \nonumber \\
     &+&\frac{1}{2}  (\partial_{\mu} \sigma \partial^{\mu}\sigma - m_{\sigma}^2 {\sigma}^2) 
     -\frac{1}{3}b m   (g_{\sigma} \sigma)^3+\frac{1}{4}c   (g_{\sigma} \sigma)^4 \nonumber\\
     &+&\frac{1}{2}m_{\omega}^2 \omega_{\mu}\omega^{\mu}-\frac{1}{4} \omega_{\mu \nu}\omega^{\mu \nu}+ {\frac{\zeta}{4!} \left (g_{\omega}^2 \omega_{\mu} \omega^{\mu}\right)^2} \nonumber \\
     &-&\frac{1}{4}  (\vec{\rho}_{ \mu \nu}.\vec{\rho}^{\mu \nu}-2 m_{\rho}^2 \vec{\rho}_{\mu}\vec{\rho}^{\mu})+\Lambda_{\omega} (g_{\rho}^2  \vec{\rho}_{\mu} \vec{\rho}^{\mu}) \  (g_{\omega}^2 \omega_{\mu}\omega^{\mu}) \nonumber \\
     &+& \sum_{l=e^-,\mu^-} \bar{\psi}_l (i \gamma_{\mu} \partial^{\mu} - m_l) \psi_l~, 
     \label{eqn:lagr}
\end{eqnarray}
where $\Psi_{N}$ is the Dirac field of the nucleons $N$, $\psi_l$ is the Dirac field for leptons, while $\gamma^{\mu}$ and $\vec{\tau}$ are the Dirac and Pauli matrices respectively. $\sigma$, $\omega$, $\rho$ denote the meson fields, with isoscalar coupling constants  $g_{\sigma}$, $g_{\omega}$, isovector coupling $g_{\rho}$ and mixed $\omega - \rho$ coupling $\Lambda_{\omega}$. $b$ and $c$ represent the scalar meson self interaction, while the coupling $\zeta$ represents the quartic vector self-interaction. The vacuum nucleon mass is $m$ while $m_l$ denotes lepton masses. The field tensors $\omega_{\mu \nu}$ and $\vec{\rho_{\mu \nu}}$ are defined as:
$$\omega_{\mu \nu} = \partial_{\mu} \omega_{\nu} - \partial_{\nu} \omega_{\mu}~,$$
$$\vec{\rho}_{\mu \nu} = \partial_{\mu} \vec{\rho}_{\nu} - \partial_{\nu} \vec{\rho}_{\mu}~.$$

\indent From the above Lagrangian density, the equations of motion for the nucleons and mesons can be obtained in the mean-field limit. The energy density can be derived from the energy-momentum tensor as,

\begin{eqnarray}
\varepsilon &=&  \sum_{N}\frac{1}{8\pi^{2}}\Big[k_{F_N}E_{F_N}^{3} + k_{F_N}^{3}E_{F_N} - {m^*}^4\ln{\dfrac{k_{F_N} + E_{F_N}}{m^*}}\Big] \nonumber\\
&+&\frac{1}{2}m_{\sigma}^{2} \sigma^{2} + \frac{1}{3}b m(g_{\sigma} \sigma)^{3}+\frac{1}{4}c(g_{\sigma} \sigma)^{4}\nonumber \\
&+& \frac{1}{2}m_{\omega}^{2} \omega^{2}+\frac{\zeta}{8}(g_{\omega} \omega)^{4}~\nonumber\\
&+& \frac{1}{2}m_{\rho}^{2} \rho^{2} + 3\Lambda_{\omega}(g_{\rho}g_{\omega} \rho \omega)^{2}~
\end{eqnarray}

The pressure can be obtained from the energy density $\varepsilon$ via the Gibbs-Duhem relation,
\begin{equation}
    P = \sum_{N}\mu_{N}n_{N} - \varepsilon
    \label{eqn:eos}
\end{equation}

where the chemical potential of nucleons is given by
\begin{equation*}
    \mu_{N} = \sqrt{k_{F_N}^2+{m^*}^2} + g_{\omega} \omega + \frac{g_{\rho}}{2}\tau_{3N} \rho~.
\end{equation*}

Mesonic equations of motion as evaluated by using Euler-Lagrange equations,

\begin{eqnarray}
    m_{\sigma}^2 \sigma &=& \sum_B g_{\sigma} n_B^s -\frac{\partial{U_{\sigma}}}{\partial \sigma} \\
    m_{\omega}^2 \omega +\frac{\zeta}{3!}g_{\omega}^4 \omega^3&=& \sum_B g_{\omega} n_B-2\Lambda_{\omega} g_{\rho}^2 g_{\omega}^2 \rho^2 \omega \\
m_{\rho}^2 \rho =& \sum_B&  g_{\rho} I_{3_B}n_B-2\Lambda_{\omega} g_{\rho}^2 g_{\omega}^2 \omega^2 \rho
\end{eqnarray}
\label{eqn:mesonfieldeq}
where $n_B^s$ and $n_B$ are scalar and vector  baryon densities respectively.\\

The isoscalar set of couplings $g_{\sigma}$, $g_{\omega}$, $b$ and $c$ are determined by fixing the saturation density $n_0$, binding energy per nucleon $E_{sat}$, the incompressibility coefficient $K_{sat}$ and the effective nucleon mass $m^*= m - g_{\sigma}\sigma$ at saturation. On the other hand, isovector couplings $g_{\rho}$ and $\Lambda_{\omega}$ are determined as a function of the symmetry energy $J_{sym}$ and the slope of the symmetry energy at saturation $L$. However, the parameter $\zeta$ is fixed such that the EoS should be able to reproduce the maximum observed neutron star mass. In many works, the vector self-interaction is either ignored ~\cite{Hornick,Ghosh2022,Ghosh2022b,Pradhan2022} or they are fixed to constant value ~\cite{Chen2014,Thakur} by choosing a particular EoS model. In this work, we will consider the variation of all nuclear saturation parameters along with $\zeta$, and the ranges for each of the variables are provided in~\Cref{tab:HTZCS_parameter_set}.  As the change of the crust EoS does not affect the NS bulk properties significantly~\cite{Lattimer_2001}, we fix the crust EoS to that of ~\cite{HEMPEL2010210} and stitch the core EoS such that the EoS is thermodynamic stable ($dp/d\epsilon >0$) and also satisfies the causality requirements (i.e, the in-medium sound speed is less than the speed of light).

%\begin{figure}[h]
%    \centering
%    \caption{EoS with varying $\zeta$ values}
%   \label{fig:eos_zetas}
%\end{figure}

%\subsection{Model parameters}
%\label{sec:para}

%\dc{Tables}
 \begin{table*}
 
        \caption{Chosen parameter set used in this study. Masses of mesons are set to $m_{\sigma}=550$ MeV, $m_{\omega}=783$ MeV, $m_{\rho}=770$ MeV and nucleon mass is fixed to $m_N=939$ MeV. }\label{tab:HTZCS_parameter_set}
        \centering\small\setlength\tabcolsep{.33em}
        \begin{tabular}{ c  c  c  c c  c  c  c }
        \hline \hline
       Model& $n_0$ $\rm (fm^{-3})$& $E_{sat}$ (MeV) & $K_{sat}$  (MeV) & $J_{sym}$ (MeV) & $L$ (MeV) & $m^*/m_N$ & $\zeta$ \\
        \hline
         Hornick et al.~\cite{Hornick}  & 0.150 & -16.0 & 240 & 32 & 60 & 0.65 & 0.00\\
       
        Variation & [0.14, 0.17] & [-16.5, -15.5] & [200, 300] & [28, 34] & [40, 70] & [0.55, 0.75] & [0.00, 0.1]\\
        \hline \hline
        \end{tabular}
        
        \end{table*}
%%%%%%%%%%%%%%%%%%%%%%%%%%%%%%%%%%%%%%%%

\subsection{Macroscopic NS properties}

%\dc{Mass, Radius, Tidal deformability using TOV equations}
Given an EoS, one can calculate the mass-radius (M-R) relationship of non-rotating NSs using the Tolman-Oppenheimer-Volkoff (TOV) equations of hydrostatic equilibrium 

\begin{eqnarray}
    \frac{dm(r)}{dr} &=& 4\pi \epsilon(r)r^{2}, \nonumber\\
    \frac{dp(r)}{dr} &=& -\frac{[p(r) + \epsilon(r)][m(r) + 4\pi r^{3}p(r)]}{r(r - 2m(r))}~.
    \label{eq:MR_relation}
\end{eqnarray}

\indent For a given EoS (pressure-density relation), stable  stellar configurations are obtained by integrating  TOV Eqs~\eqref{eq:MR_relation}  from the center to the surface of the star with the proper boundary condition that the pressure vanishes at the surface, i.e, $p(R)=0$ to obtain the radius $R$ and the enclosed mass within $R$ presents the NS mass $M=m(R)$.\\

\indent In an NS merger, coalescing binary NSs get deformed due to highly distorted space-time surrounding them (acts as an external tidal field) and this is measured as tidal deformability. The  tidal deformability depends upon the NS composition and can be used to constrain the NS EoS. One can derive important information about the structure of NSs from the tidal deformability which is given as a function of radius $R$ and tidal love number $k_{2}$ in Eq.~\eqref{eq:tidaldef}. The love number $k_{2}$ can be obtained by solving an additional set of differential equations along with the TOV equations ~\cite{Hinderer2008}.

\begin{eqnarray}
    \lambda = \frac{2}{3}k_{2}R^{5}
    \label{eq:tidaldef}
\end{eqnarray}
The dimensionless tidal deformability is defined as,
\begin{eqnarray}
    \bar{\Lambda}= \frac{2}{3}k_{2}C^{-5}
    \label{eq:diml_tidaldef}
\end{eqnarray}
where $C$ is the compactness i.e., the ratio of mass and radius $C = M/R $.

\section{Results}
\label{sec:results}

\subsection{Preliminary studies}
\label{sec:prelim}

%without zeta and with zeta comparison, of fields, eos, MR?

\par
\begin{figure*}
    %\centering % Not needed
    \begin{subfigure}[b]{0.45\textwidth}
        \includegraphics[width=\textwidth]{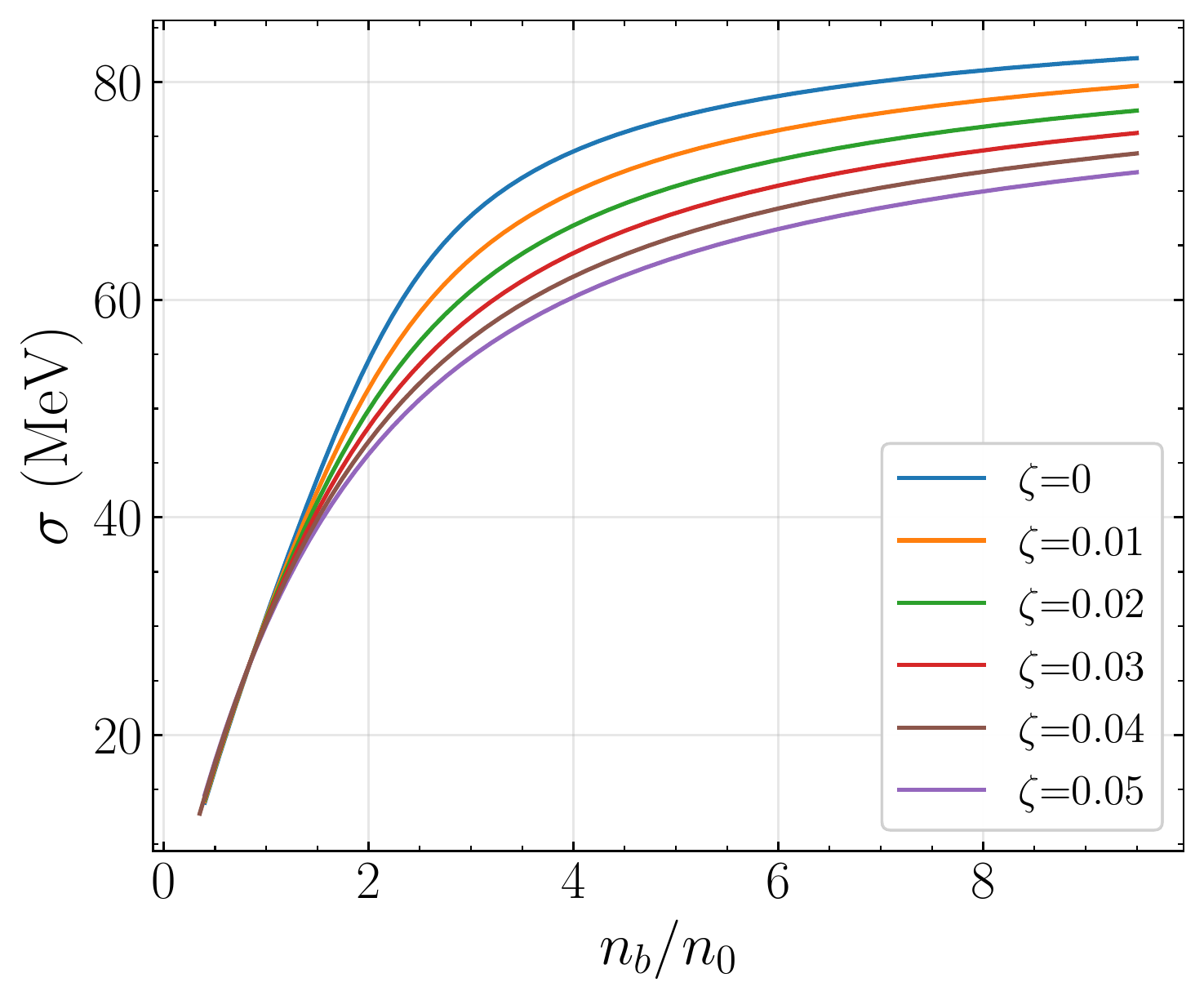}
        \caption{ }
        \label{subfig:sigma_field}
    \end{subfigure}
    \hfill
    \begin{subfigure}[b]{0.45\textwidth}
        \includegraphics[width=\textwidth]{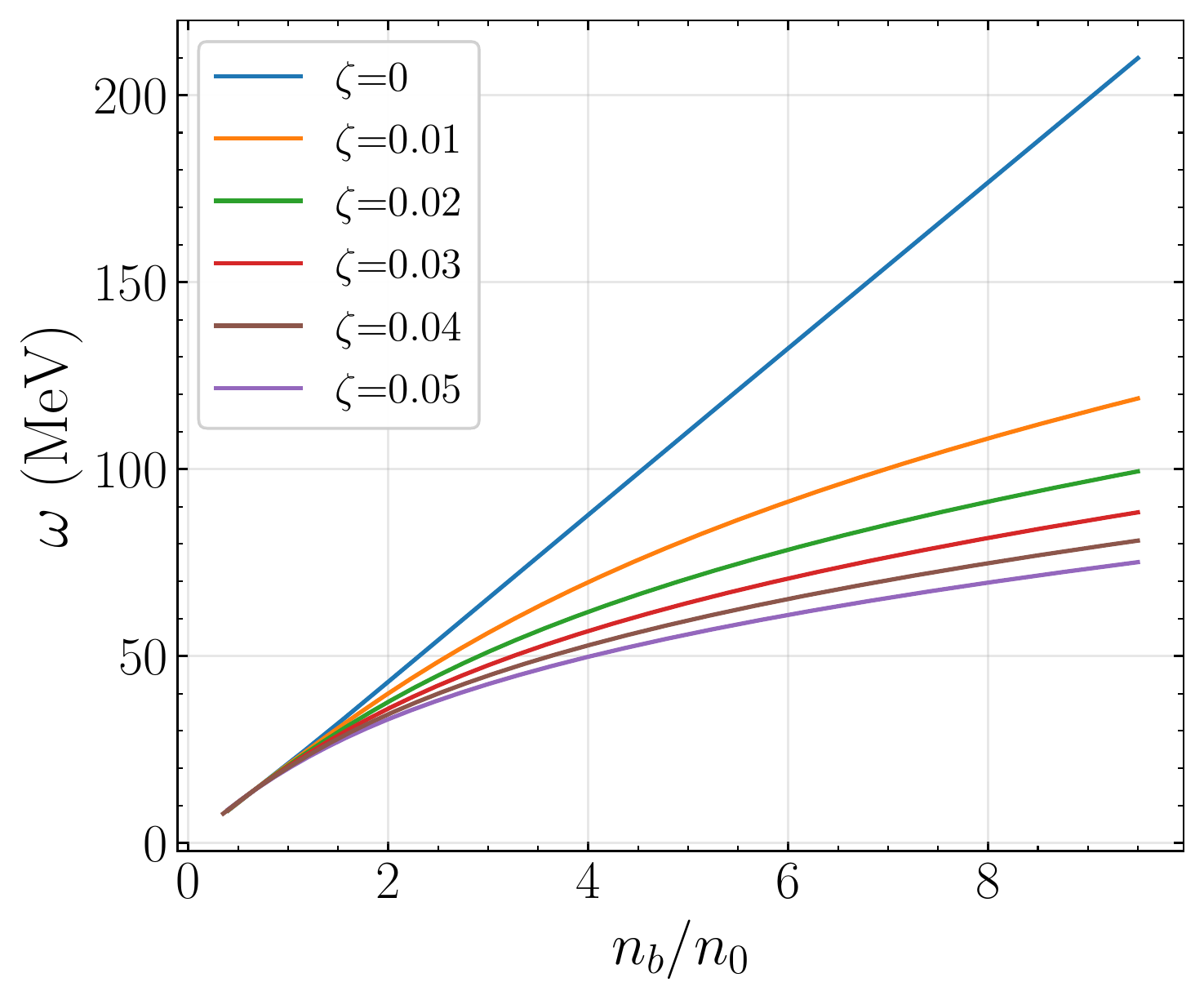}
        \caption{}
        \label{subfig:omega_field}
    \end{subfigure}
    %% leave a blank line to create a line break

    \begin{subfigure}[b]{0.45\textwidth}
        \includegraphics[width=\textwidth]{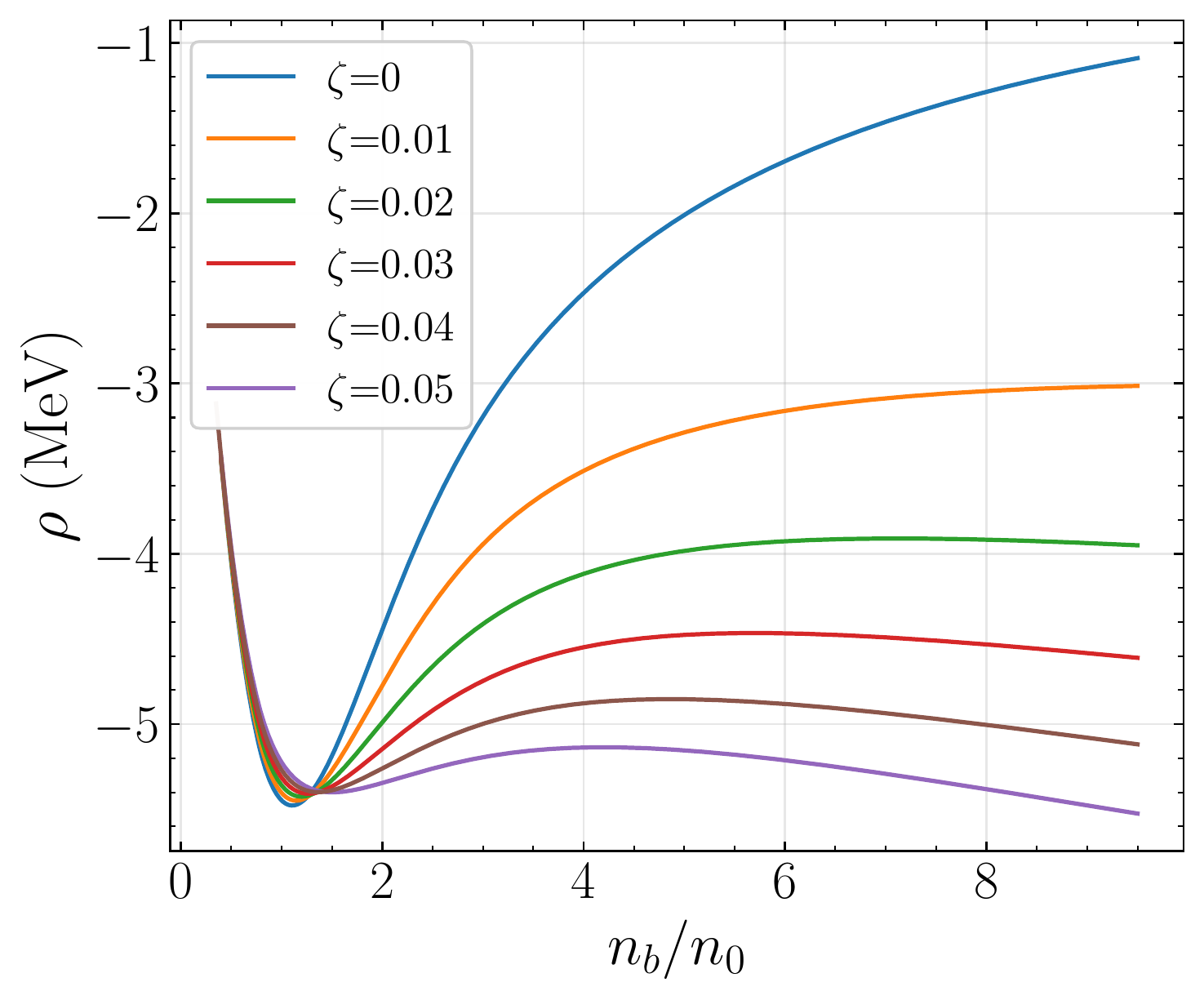}
        \caption{}
        \label{subfig:rho_field}
    \end{subfigure}
    \hfill
    \begin{subfigure}[b]{0.45\textwidth}
        \includegraphics[width=\textwidth]{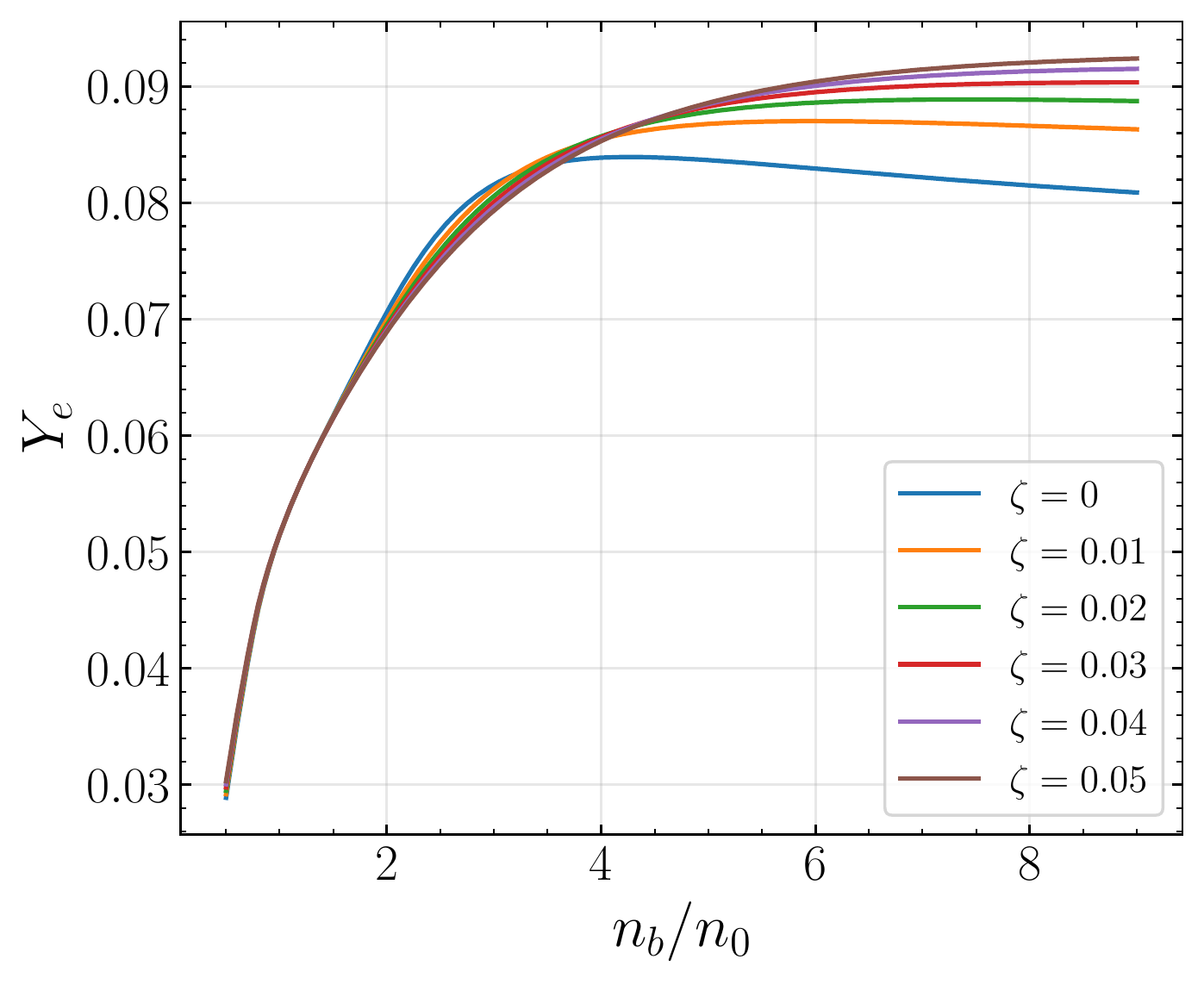}
        \caption{}
        \label{subfig:efrac}
    \end{subfigure}
    \caption{Mean meson field strengths of neutron star matter as a function of baryon number density with different $\zeta$ values, (a) for isoscalar $\sigma$ meson,(b) for isoscalar-vector $\omega$ meson and (c) for isovector $\rho$ meson. (d) Shows the variation of the electron fraction $Y_e$ in the NS core as a function of baryon number density for different $\zeta$ values.}
    \label{fig:meson_fields}
\end{figure*}

 Before a systematic investigation of the impact of $\zeta$,  we perform a preliminary study where we fix other nuclear saturation parameters to the parameter set for Hornick et al.~\cite{Hornick} as  given in~\Cref{tab:HTZCS_parameter_set} and vary the values of $\zeta$. Modified coupling constants, as well as the corresponding changes in the properties of a canonical 1.4$M_{\odot}$ NS obtained for increasing $\zeta$, are tabulated in \Cref{tab:NS_at_diff_zeta}. As expected from previous works~\cite{Horowitz2001,MULLER1996}, the EoS softens with increasing contribution of the vector self-interaction $\zeta$ and hence reduces the maximum mass that can be supported by the EoS. We further investigate the behavior of meson fields as a function of density  with different vector self-interaction strength and display the behavior of  $\sigma$, $\omega$ and $\rho$ meson fields in ~\Cref{subfig:sigma_field,subfig:omega_field,subfig:rho_field} respectively. Additionally, we display the variation of electron fraction ($Y_e$, see ~Eq.~\eqref{eqn:efrac} for definition) of the NS matter with different EoS models in ~\Cref{subfig:efrac}. It was discussed in earlier works~\cite{Horowitz2001,MULLER1996}, that the vector meson field $V = g_\omega \omega$ is related to the baryon density $n_b$ through the relation~\eqref{eqn:omega_field}.
From ~\Cref{subfig:omega_field}, we observe that the linear behavior of omega meson field with baryon density ($n_b$) at $\zeta=0$  changes to the relation given in ~Eq.\eqref{eqn:omega_field} for finite non zero $\zeta$ values ~\cite{MULLER1996}, i.e.,  a finite $\zeta$ leads
to an effective medium dependence of the $\omega$ mass~\cite{Ryszard_2001}.

 \begin{eqnarray}
 V \left(\frac{m_{\omega}^2}{g_{\omega}^2} + \frac{\zeta}{6} V^2 \right) = n_b~.
    \label{eqn:omega_field}
\end{eqnarray}

For the case $\zeta$ = 0,  it reduces to 
\begin{eqnarray}
    V = \frac{g_{\omega}^2}{m_{\omega}^2} n_b
\end{eqnarray}

The electron fraction `$Y_e$' is defined as,
\begin{equation}\label{eqn:efrac}
    Y_e=n_e/n_b
\end{equation}
where $n_e$ is the electron number density and $n_b$ is the baryon number density.

\begin{figure}[htbp]
    \centering
     \begin{subfigure}{0.325\textwidth}
      \includegraphics[width=\linewidth]{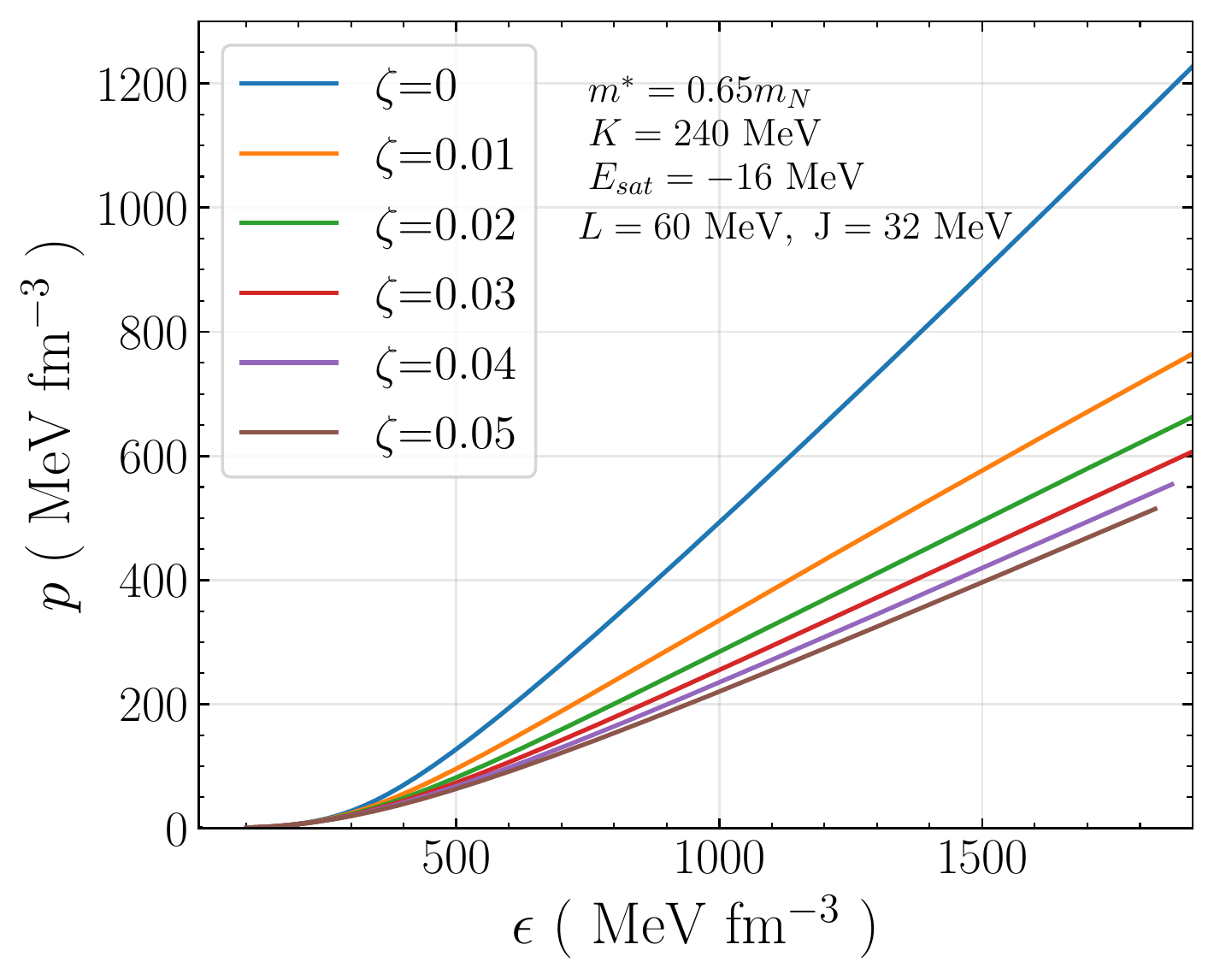}
      \caption{}
      \label{subfig:Eos_diff_zeta}
    \end{subfigure}
    \begin{subfigure}{0.325\textwidth}
      \includegraphics[width=\linewidth]{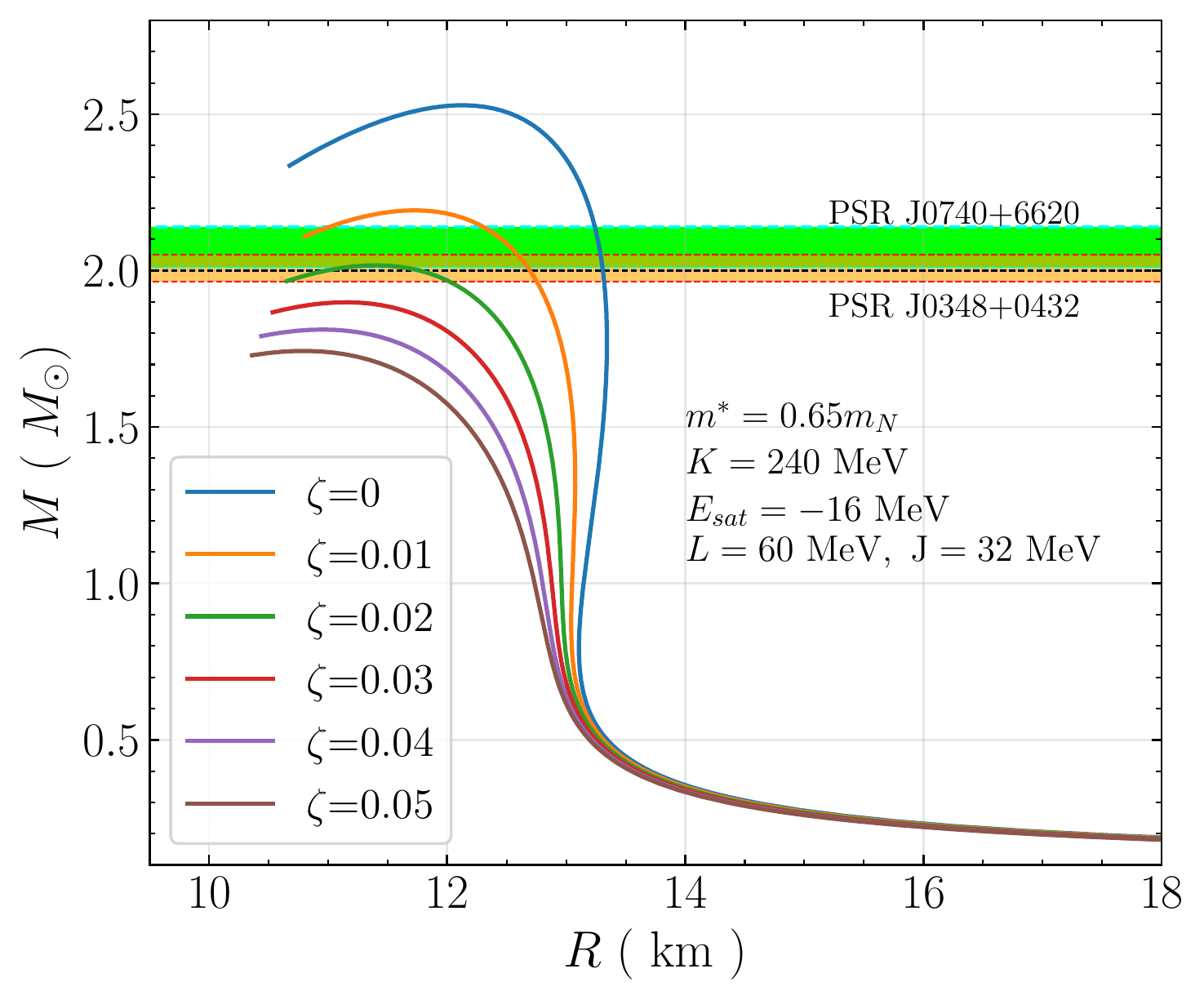}
      \caption{}
      \label{subfig:MR_diff_zeta}
    \end{subfigure}
    \begin{subfigure}{0.325\textwidth}
      \includegraphics[width=\linewidth]{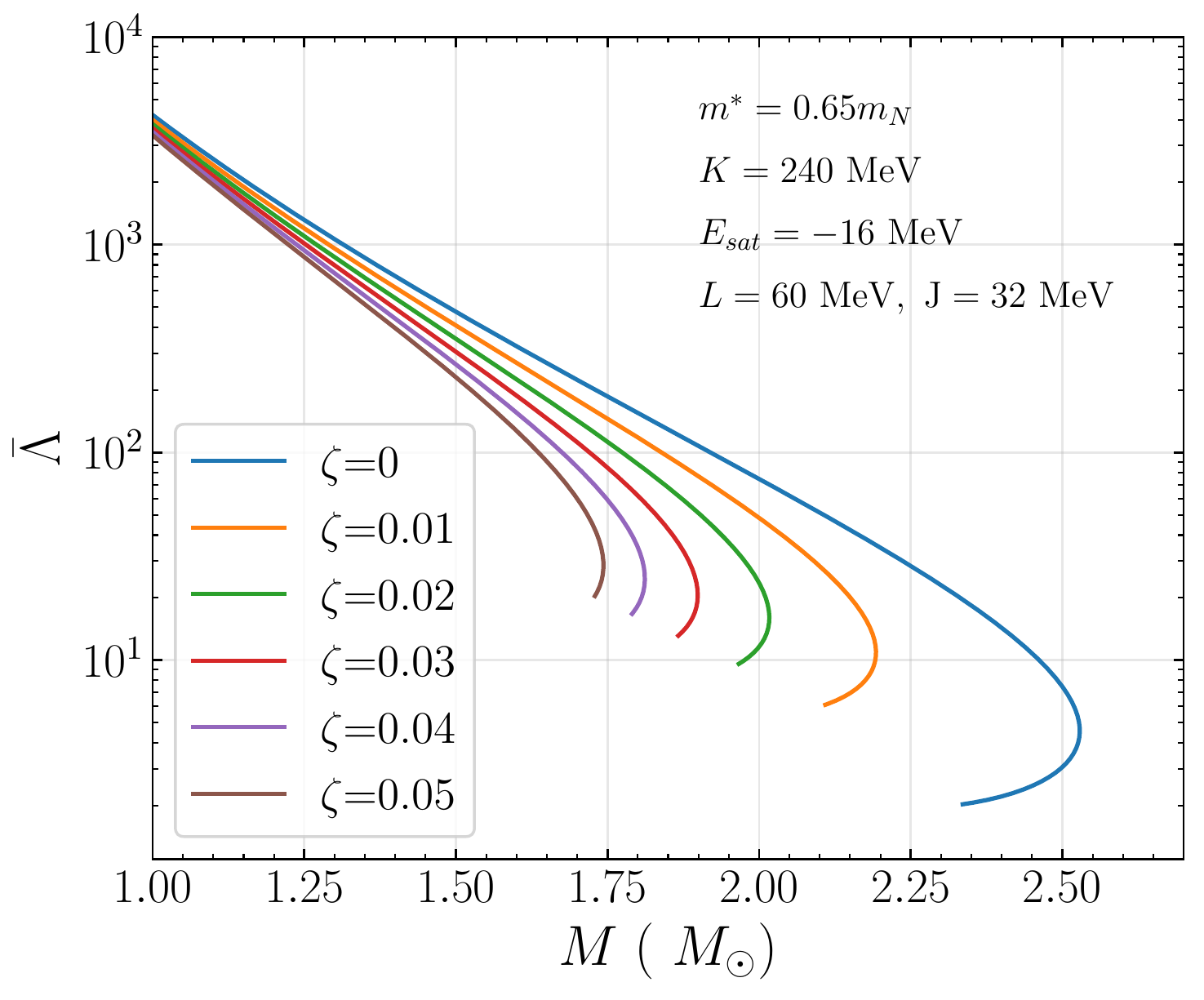}
      \caption{}
      \label{subfig:tidal_diff_zeta}
    \end{subfigure}
    \caption{(a) Equations of state with different $\zeta$ values. (b) Mass-radius relations and (c) dimensionless tidal deformability ($\bar{\Lambda}$) are displayed as a function of stellar mass corresponding to EoSs displayed in~\Cref{subfig:Eos_diff_zeta}.}
 \label{fig:eos_mr}
\end{figure}

The EoS, mass-radius relations and tidal deformability corresponding to the nuclear parameter set and different values of $\zeta$ given in~\Cref{tab:NS_at_diff_zeta} are displayed in~\Cref{subfig:Eos_diff_zeta,subfig:MR_diff_zeta,subfig:tidal_diff_zeta} respectively. The low density behavior of nuclear matter is  ($0.5 \leq n_b/n_0\leq 1.5$) constrained by the properties of the pure neutron matter (PNM) resulting from  recent  chiral effective field theory ($\chi \rm EFT$)~\cite{Drischler}. To impose the $\chi \rm EFT$ constraints, we check whether or not the binding energy of the PNM matter at each density lies within the limiting  values resulting from~\cite{Drischler}. The EoS model satisfies the astrophysical constraints, if it satisfies the following, (1) the model is able to produce a stable $2M_{\odot}$ NS (2) the reduced tidal deformability ($\tilde{\Lambda}$ ~\cite{Abbott2019}) for the binary companion of  GW170817 is  $\leq 800$~\cite{Abbott2018,Abbott2019} (this can be loosely implemented using $\bar{\Lambda}_{1.4M_{\odot}}\leq 800$) and (3) the EoS model should also produce the radii within the maximum  limit resulting for electromagnetic observations  at a different region of NS masses resulting from NICER measurements ~\cite{Miller_2019,Miller_2021}. For a given EoS model, in ~\Cref{tab:NS_at_diff_zeta} we provide a check mark (cross mark) if the model satisfies (does not satisfy) the astrophysical and $\chi \rm EFT$ constraint. The models listed in~\Cref{tab:NS_at_diff_zeta} all satisfy the $\chi \rm EFT$ constraints. At such low density $\zeta$ has no significant impact on the meson fields~(see,~\Cref{fig:meson_fields} ) or on the  PNM properties~ (see~\Cref{fig:PNM_diff_zeta} in Appendix)~\cite{Chen2014,Horowitz2001,Dhiman2007}. However, increasing $\zeta$ from 0.03, the EoS models fail to produce  the maximum $2M_{\odot}$ NS. This implies that, although the impact of the vector self-interaction at low density is insignificant, $\zeta$ controls the high-density behavior of dense matter, and therefore astrophysical constraints at high density can be used to constrain its maximum value (in fact in~\Cref{subsec:limit_zeta} we will see that $\zeta$ should be $\leq 0.033$ to satisfy the astrophysical constraints).

 EoS tables for  different parameterized models tabulated in ~\Cref{tab:NS_at_diff_zeta} are provided in a public  GitHub repository~\footnote{\url{https://github.com/bikramp-hub/PCSB_EoSs}} and can be  used for future works.
%\dc{Did not follow this comment}\textcolor{black}{ [response:  I wanted to mention that the $\zeta$ mostly controls the high density behaviour of NS matter and not the low density behaviour, so $\zeta$ can be constrained from Astrophysical observations and not the constraints like $\chi \rm EFT$ ]} 
%\dc{Add comment about whether results compatible with NICER?}\textcolor{black}{Response:I have added in the paragraph that ``(3) the EoS model should also produce the radii within the maximum  limit resulting for electromagnetic observations  at different region of NS masses  ~\cite{Miller_2019,Miller_2021}." The cited articles are from NICER and also I modified the line a little.}

%\rg{Should we keep zeta = 0.05 coupling constants in the following table?}
%\bkp{For preliminary results we can keep it. }

\begin{table*}[htbp]
\caption{Coupling constants, radius and  tidal deformability  of a $1.4M_{\odot}$ neutron star (NS) ($R_{1.4M_{\odot}}$ and $\bar{\Lambda}_{1.4M_{\odot}}$) and the maximum possible NS mass for different $\zeta$ values are tabulated. Other nuclear saturation parameters are fixed to the values given for Hornick et al.  in \Cref{tab:HTZCS_parameter_set}.}
    \label{tab:NS_at_diff_zeta}
    \centering\small\setlength\tabcolsep{.33em}
    \begin{tabular}{c c c c c c c c c c c c c}
    \hline \hline
    Model&$\zeta$ &$g_{\sigma}$ &$g_{\omega}$& $g_{\rho} $&$b$&$c$&$\Lambda_{\omega}$ &$R_{1.4M_{\odot}}$ &$\bar{\Lambda}_{1.4M_{\odot}}$&$M_{\rm max} $ & Astro & $\chi \rm EFT$ \\
    & & & & & & & & (km) &  & $(M_{\odot})$ \\
    \hline
    PCSB0 &0&10.429 & 11.774 & 10.273 &  0.003084  & -0.003682 & 0.027841
    &13.487 & 707.904 & 2.53&\checkmark&\checkmark\\
    
    PCSB1&0.01&10.488 & 11.928 & 10.357  & 0.002694 & -0.002280 & 0.028508&13.262 & 623.650 & 2.19&\checkmark&\checkmark\\
    
   PCSB2& 0.02 &10.548 & 12.088 & 10.443 & 0.002308 & -0.000886 &0.029175 & 13.051 & 553.791 & 2.02&\checkmark&\checkmark\\
    
    PCSB3&0.03 &10.611 &12.254 & 10.531 & 0.001927& 0.000499 &0.029842&12.852 & 494.164 & 1.89&\xmark&\checkmark\\

    PCSB4&0.04 & 10.675 & 12.427& 10.620&  0.001550& 0.001875& 0.030509&12.524 & 442.134 & 1.81&\xmark&\checkmark\\
    
    PCSB5&0.05 &10.742 & 12.608 & 10.714 & 0.001178 &0.003242& 0.031176&12.476 & 397.598 & 1.74&\xmark&\checkmark\\
   \hline \hline
    \end{tabular}
    
\end{table*}
\begin{comment}

\begin{figure}[htbp]
  \centering
  
  \begin{minipage}{.47\linewidth}
    \centering
    \begin{subfigure}{\textwidth}
      \includegraphics[width=\linewidth]{sigma_k240m0.65j32l60.pdf}
      \caption{}
    \end{subfigure}
    \begin{subfigure}{\textwidth}
      \includegraphics[width=\linewidth]{omega_k240m0.65j32l60.pdf}
      \caption{}
    \end{subfigure}
    \begin{subfigure}{\textwidth}
      \includegraphics[width=\linewidth]{rho_k240m0.65j32l60.pdf}
      \caption{}
    \end{subfigure}
   % \subcaptionbox{}{\includegraphics[width=\linewidth]{sigma_k240m0.65j32l60.pdf}}

   % \subcaptionbox{}{\includegraphics[width=\linewidth]{sigma_k240m0.65j32l60.pdf}}

    %\subcaptionbox{}{\includegraphics[width=\linewidth]{sigma_k240m0.65j32l60.pdf}}
    
  \end{minipage}
  \hfill
  \begin{minipage}{.47\linewidth}
    \centering
    \subcaptionbox{}{\includegraphics[width=\linewidth]{eos_k240m0.65j32l60.pdf}}

    \subcaptionbox{}{\includegraphics[width=\linewidth]{mr_k240m0.65j32l60.pdf}}

    \subcaptionbox{}
      {\includegraphics[width=\linewidth]{mlambda_k240m0.65j32l60.pdf}}

    %
  \end{minipage}
   \caption{Mean meson field strengths as a function of density with different $\zeta$ values, (a) for isoscalar $\sigma$ meson,(b) for isoscalar-vector $\omega$ meson and (c) for isovector $\rho$ meson.}
\end{figure}

\end{comment}

\begin{comment}

\begin{figure}[htp]

\begin{subfigure}{\linewidth}
\includegraphics[clip,width=\linewidth]{sigma_k240m0.65j32l60.pdf}
\caption{}
\label{subfig:sigma_field}
\end{subfigure}

\begin{subfigure}{\linewidth}
\includegraphics[clip,width=\linewidth]{omega_k240m0.65j32l60.pdf}
\caption{}
\label{subfig:omega_field}
\end{subfigure}
 
\begin{subfigure}{\linewidth}
\includegraphics[clip,width=\linewidth]{rho_k240m0.65j32l60.pdf}
\caption{}
\label{subfig:rho_field}
\end{subfigure}
\caption{Mean meson field strengths as a function of density with different $\zeta$ values, (a) for isoscalar $\sigma$ meson,(b) for isoscalar-vector $\omega$ meson and (c) for isovector $\rho$ meson.}
  \label{fig:meson_fields}

\end{figure}

\begin{figure}[htp]

\begin{subfigure}{\linewidth}
\includegraphics[clip,width=\linewidth]{eos_k240m0.65j32l60.pdf}
\caption{}
\label{subfig:Eos_diff_zeta}
\end{subfigure}

\begin{subfigure}{\linewidth}
\includegraphics[clip,width=\linewidth]{mr_k240m0.65j32l60.pdf}
\caption{}
\label{subfig:MR_diff_zeta}
\end{subfigure}
 
\begin{subfigure}{\linewidth}
\includegraphics[clip,width=\linewidth]{mlambda_k240m0.65j32l60.pdf}
\caption{}
 \label{subfig:tidal_diff_zeta}
\end{subfigure}
\caption{(a) Equations of state with different $\zeta$ values, (b) corresponding mass-radius relations and (c) Dimensionless tidal deformability ($\bar{\Lambda}$) as a function of stellar mass corresponding to EoSs displayed in~\Cref{subfig:Eos_diff_zeta} .}
 \label{fig:eos_mr}

\end{figure}
\end{comment}

\subsection{Correlation studies}
\label{sec:correlation}
We investigate the correlations among the nuclear saturation parameters themselves as well as with NS properties. Astrophysical observations come with uncertainties in their observed values. To consider the effect of uncertainties
of astrophysical observations, we assign a statistical weighting factor ($W$) to each parameter set. Assuming a Gaussian likelihood, the statistical weight for a parameter set $\{\rm P \}$ can be defined as~\cite{Chatterjee2017,Tsang2012},

\begin{equation}
    W_{\{\rm{P}\}}=w[\{\rm{P}\}]\exp{\left(-\chi^2_{\{\rm{P}\}}/2\right)}~,
    \label{eqn:weight}
\end{equation}

where the  weighted  sum of squared deviation for a parameter set \{P\} can be defined as
 \begin{equation}
     \chi^2_{\{\rm{P}\}}=\sum_{i} \frac{(O_i-C_i[\{\rm P\}])^2}{\sigma_i^2}~.
 \end{equation}
 
For astrophysical observations, we consider the observed mean ($O$) and standard deviation ($\sigma$) corresponding to mass measurement of PSR J0740+6620 ~\cite{Riley_2021} ($2.072^{+0.067}_{-0.066}$) and the tidal deformability of a $1.4M_{\odot}$ NS resulting from GW170817 ~\cite{Abbott2018} ($190^{+390}_{-120}$). $C [\rm \{P\}]$ is the calculated value obtained for a parameter set \{P\}. To a parameter set \{P\}, we assign the weight independently corresponding to each astrophysical observation and then multiply them to get total weight (i.e, $W_{\{\rm{P}\}}=W_{\{\rm{P}\},\  PSR}\times W_{\{\rm{P}\},\ \bar{\Lambda}_{1.4M_{\odot}}}$). For EoSs with maximum mass $\geq 2.072$, we assign $W_{\{\rm{P}\},\  PSR}=1$. The window function $w[\{\rm P\}]$ in Eq.~\eqref{eqn:weight} is defined such that, it has a value 1 for a parameter set \{P\}, if it satisfies the necessary conditions like the causality ( i.e, the speed of sound in the NS interior $<$ speed of light ), stability ( the pressure is a monotonic function of density) and the EoS must able to produce a stable $2M_{\odot}$ NS (additionally we have also implemented that the tidal deformability of a canonical 1.4$M_{\odot}$ is within the upper limit (800) resulting from GW170817~\cite{Abbott2019}.)~. The weighted Pearson's correlation coefficient between two variables `$X$' and `$Y$' in presence of weight vector $W$ (corr($X,Y;W$)) can be defined as,

\begin{equation}
    \rm{corr}(X,Y;W)=\frac{\rm{cov}(X,Y;W)}{\sqrt{\rm{cov}(X,X;W) \ \rm{cov}(Y,Y;W)}}
    \label{eqn:correlation}
\end{equation}
where cov($X,Y;W$) is the covariance between two variables $X$ and $Y$ and defined as,
\begin{equation}
    \rm{cov}(X,Y;W)= \frac{\sum_i W_i [X_i-M(X;W)] [Y_i-M(Y;W)]}{\sum_i W_i}
\end{equation}
where $M(X;W)$ is the weighted mean of variable $X$ $\left(i.e, M(X;W)= \frac{\sum_i W_i\ X_i}{\sum_i W_i} \right)$~.

\subsubsection{Constant $\zeta$ RMF models  and correlations}
\label{subsec:corr_constant_zeta}
\begin{figure*}[htbp]
\centering
\begin{subfigure}{.45\textwidth}
  \centering
  \includegraphics[width=\linewidth,height=7.3cm]{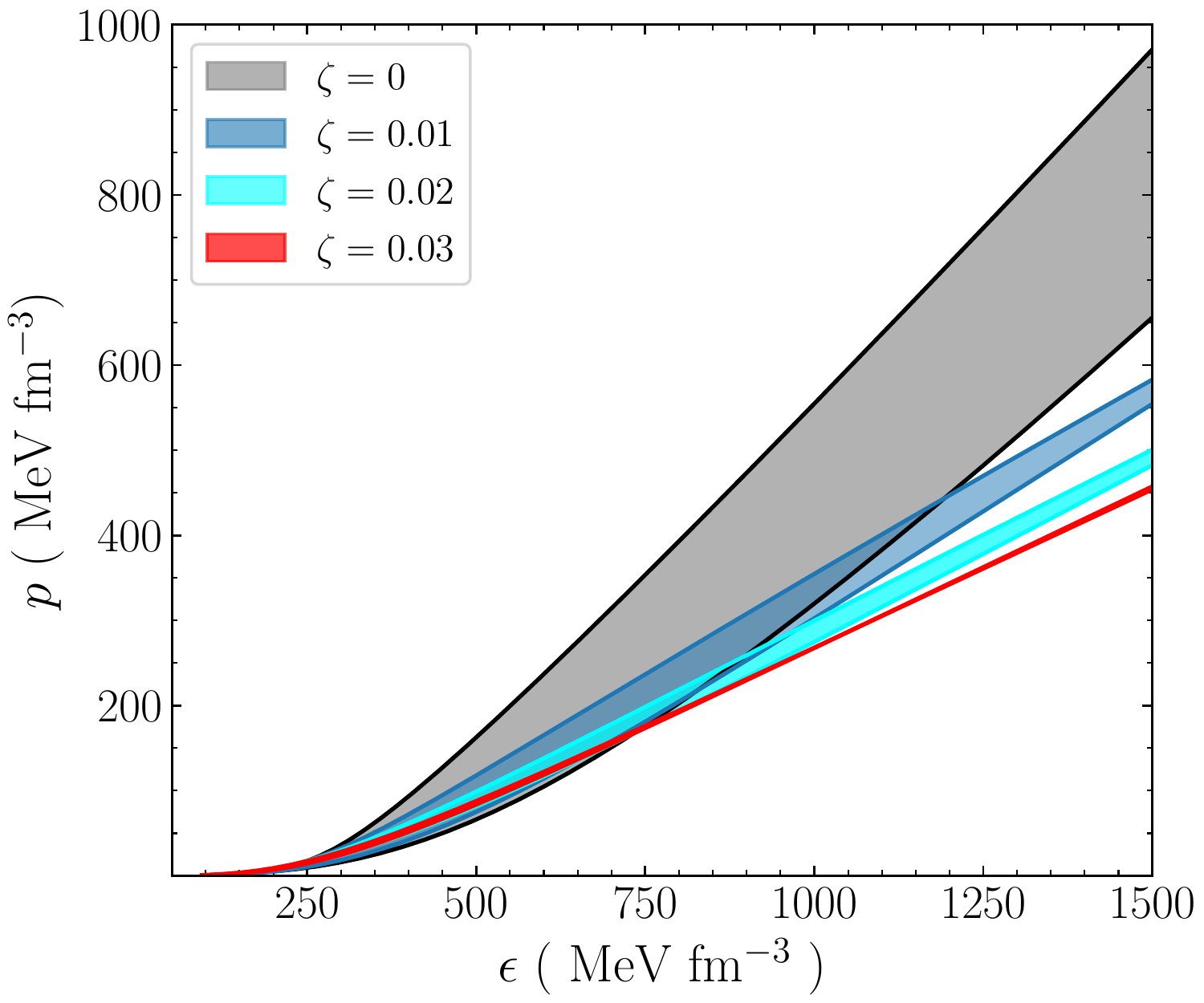}
  \caption{}  
  \label{subfig:eos_range_diffzeta}
\end{subfigure}%
\begin{subfigure}{.45\textwidth}
  \centering
  \includegraphics[width=\linewidth,height=7cm]{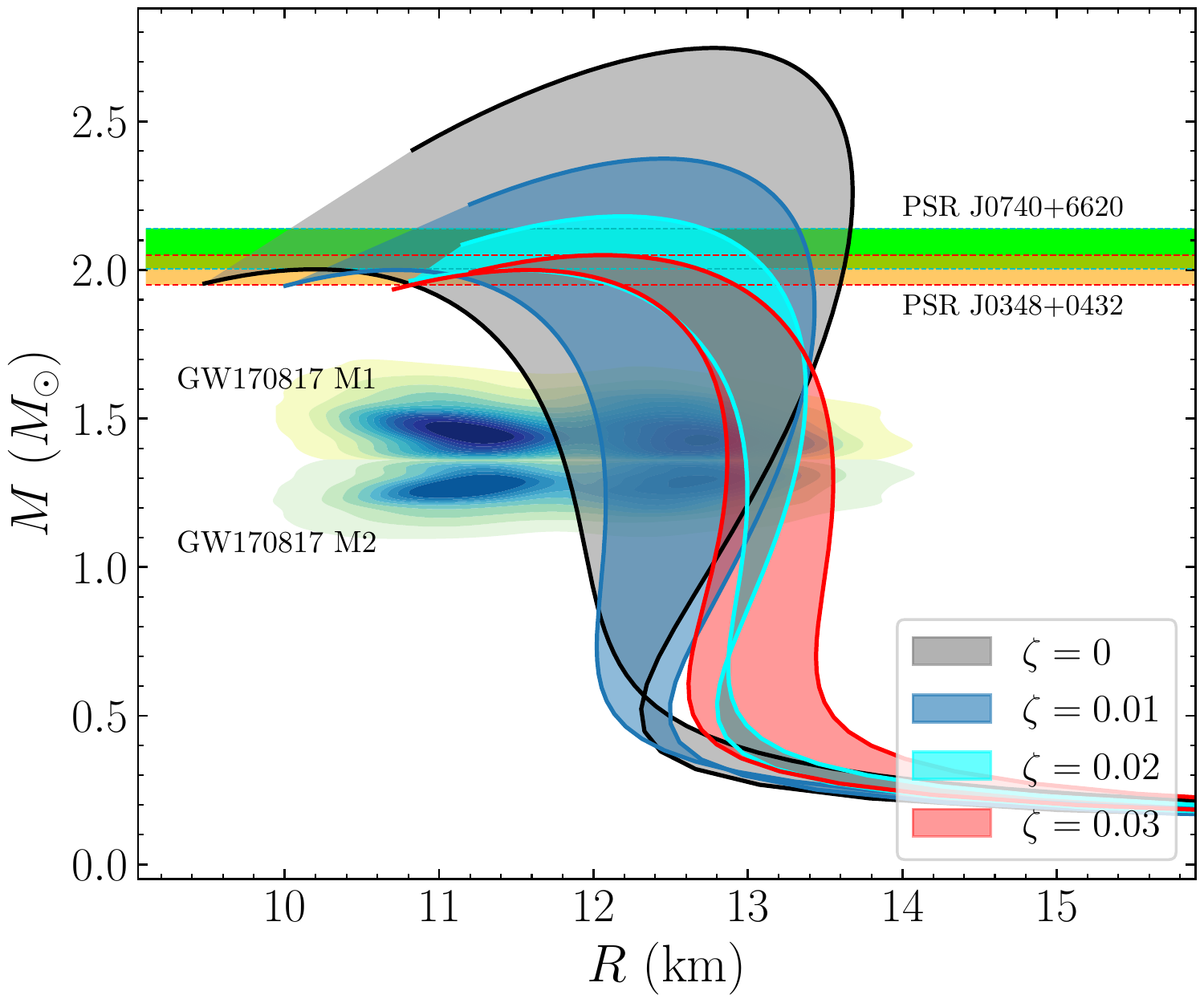}
  \caption{}
  \label{subfig:mr_range_diffzeta}
\end{subfigure}
\caption{(a) Range of EoSs with different  constant $\zeta$ models (other saturation parameters are varied uniformly) and (b) corresponding mass-radius relations. Horizontal bands correspond to masses $M=2.072^{+0.067}_{-0.066} M_{\odot}$ of PSR J0740+6620 ~\cite{Riley_2021} and $M=2.01^{+0.04}_{-0.04} M_{\odot}$ of PSR J0348$+$0432 ~\cite{Antoniadis}. The mass-radius estimates of the two companion neutron stars in the merger event GW170817~\cite{Abbott2018} are shown by the shaded area labeled with GW170817 M1 (M2).
}
\label{fig:eos_mr_range_diff_zeta}
\end{figure*}

We investigate the impact of $\zeta$ on the correlations among the nuclear saturation parameters and global NS properties for different constant $\zeta$ values. We consider $\zeta \in [0,0.03]$, as higher values of $\zeta$ give maximum masses below the highest observed 2$M_{\odot}$ (see~\Cref{tab:NS_at_diff_zeta}).The area spanned in the energy density and pressure plane corresponding to different constant $\zeta$ models subject to astrophysical constraints are displayed in~\Cref{subfig:eos_range_diffzeta} and the corresponding mass-radius relation is presented in ~\Cref{subfig:mr_range_diffzeta}~. From~\Cref{fig:eos_mr_range_diff_zeta}, one can conclude that with increasing the values of $\zeta$, the range of EoSs as well as the spread in the $M-R$ relation in the vicinity of $M\sim 2M_{\odot}$ becomes narrower. This is expected, as with increasing $\zeta$ the range of the parameter sets producing maximum stable NS mass higher than $2M_{\odot}$ becomes smaller.

\begin{figure*}[htbp]
 \begin{subfigure}{0.49\textwidth}
     \includegraphics[width=\textwidth]{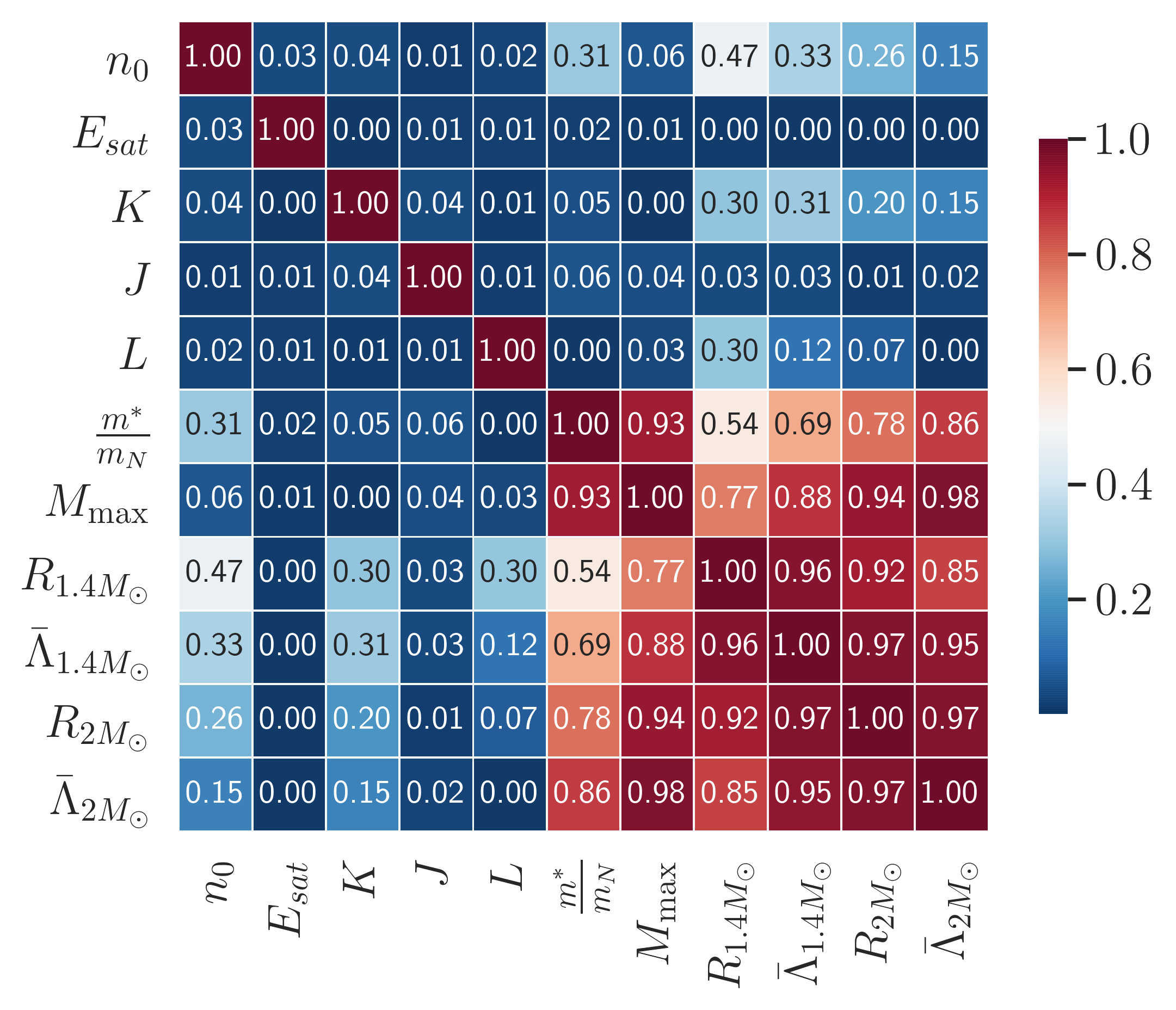}
     \caption{$\zeta=0$}
     \label{subfig:corr_zeta_0}
 \end{subfigure}
 \hfill
 \begin{subfigure}{0.49\textwidth}
     \includegraphics[width=\textwidth]{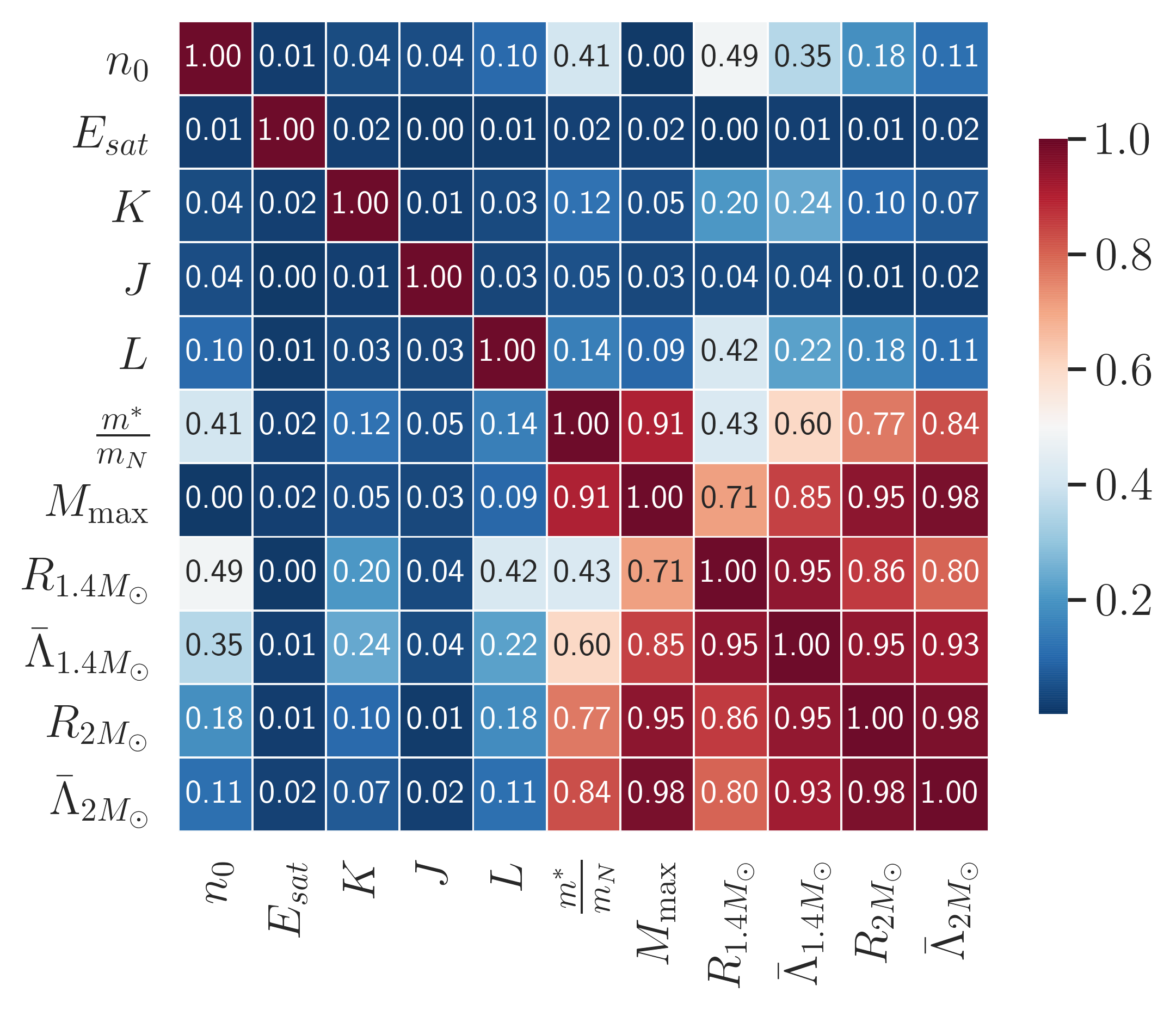}
     \caption{$\zeta=0.01$}
     \label{subfig:corr_zeta_0.01}
 \end{subfigure}
 
 \medskip
 \begin{subfigure}{0.49\textwidth}
     \includegraphics[width=\textwidth]{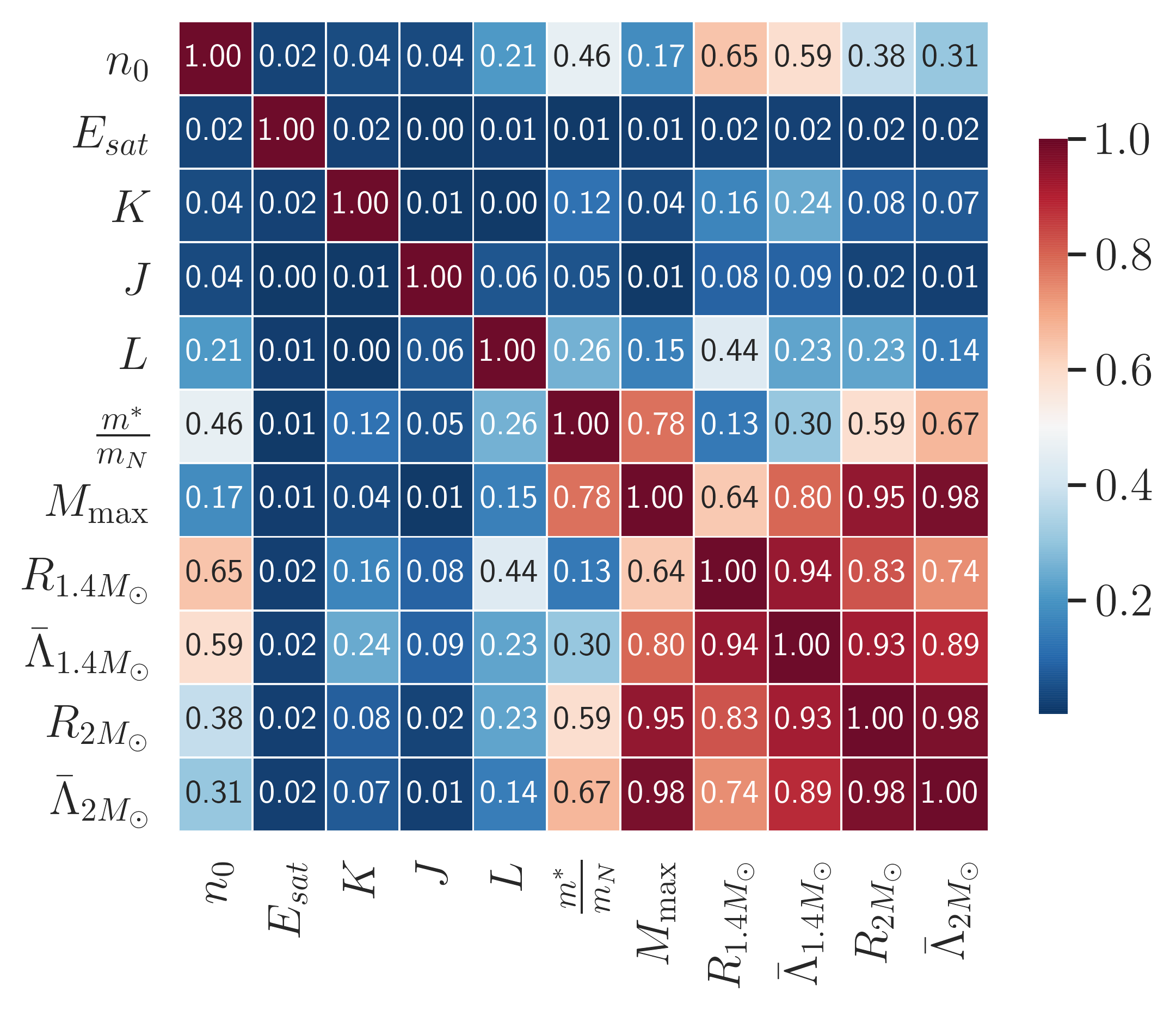}
     \caption{$\zeta=0.02$}
     \label{subfig:corr_zeta_0.02}
 \end{subfigure}
 \hfill
 \begin{subfigure}{0.49\textwidth}
     \includegraphics[width=\textwidth]{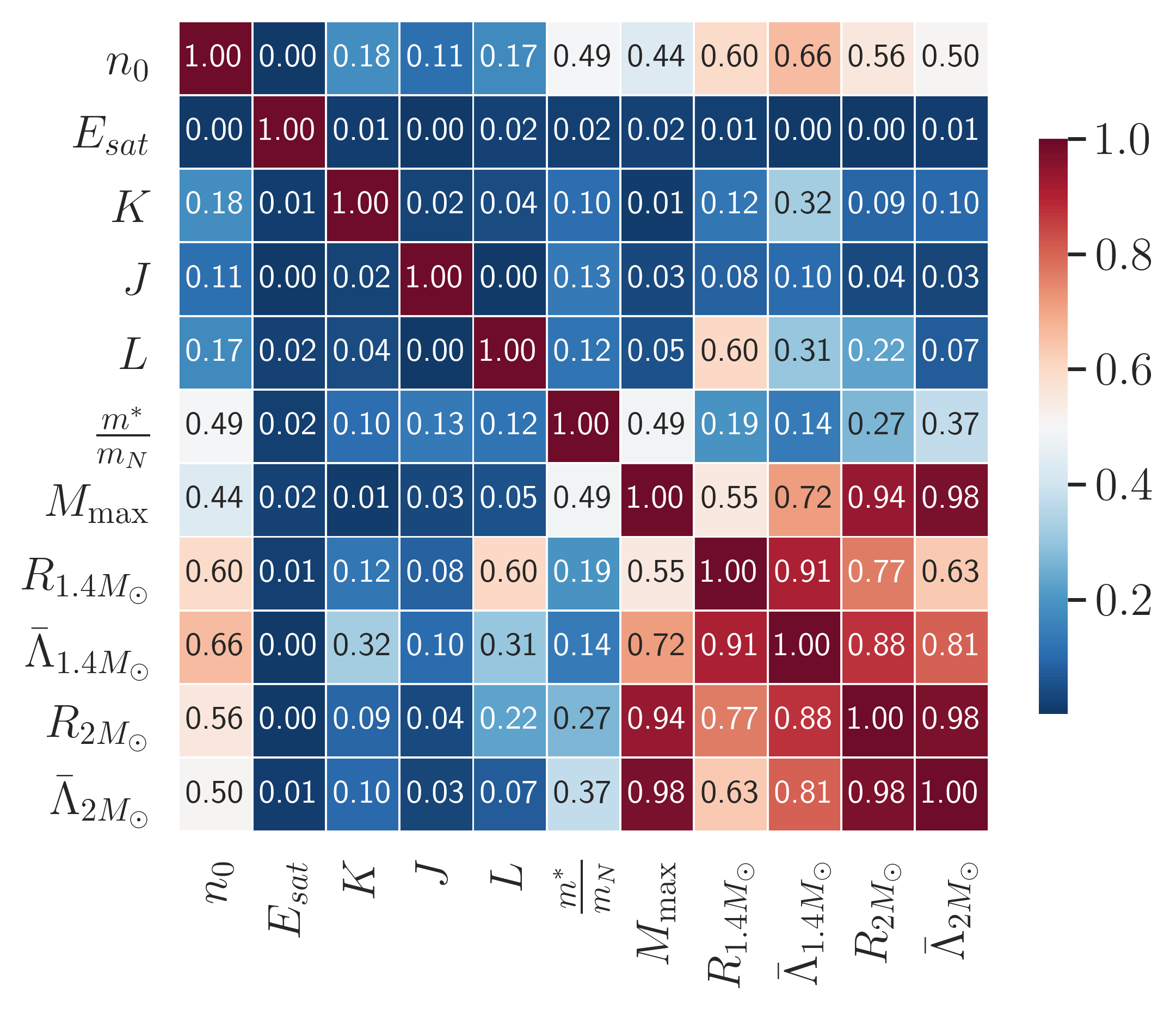}
     \caption{$\zeta=0.03$}
     \label{subfig:corr_zeta_0.03}
 \end{subfigure}

 \caption{Correlation matrices showing correlation among nuclear saturation parameters themselves as well as with NS properties. (a) Models with $\zeta=0$, (b) for $\zeta=0.01$, (c) for $\zeta=0.02$ and (d) for $\zeta=0.03$. }
 \label{fig:correlation_at_diff_zeta}
\end{figure*}

Correlation matrices  corresponding to different $\zeta$ values are displayed in~\Cref{fig:correlation_at_diff_zeta}. 
We have tabulated coefficients of important correlations between nuclear saturation parameters and NS global observables in~\Cref{tab:correlation_table} to investigate how they are affected by the choices of different $\zeta$ values. From the correlation matrices presented in  ~\Cref{subfig:corr_zeta_0,subfig:corr_zeta_0.01,subfig:corr_zeta_0.02,subfig:corr_zeta_0.03} (also from~\Cref{tab:correlation_table}), one can conclude the following:

\begin{enumerate}
    \item As expected from Eq.~\eqref{eq:diml_tidaldef}, NS radius  ($R$) shows strong correlation with $\bar{\Lambda}$. The NS properties of $1.4M_{\odot}$ and $2M_{\odot}$ show strong correlations among themselves as well as with the NS maximum mass ($M_{\rm max}$).
    \item  The correlation among symmetric nuclear energy $L$ and $R_{1.4M_{\odot}}$ increases with increasing $\zeta$ and it reaches a moderate value of 0.60 at $\zeta=0.03$.
     \item $M_{\rm max}$ correlation with $m^*$ decreases with increasing $\zeta$ (from 0.93 to 0.49).
    %For $\zeta=0$ and 0.01, $M_{\rm max}$ correlates strongly with $m^*$ (0.93 for $\zeta=0$ and 0.91 for $\zeta=0.01$). However, further increasing $\zeta$ values to 0.02 and 0.03 the correlation among $M_{\rm max}$ and $m^*$ drops to 0.78 (high) and 0.49 (moderate) respectively. 
    \item The correlation between $m^*$ and both $R_{1.4M_{\odot}}$ and $R_{2M_{\odot}}$ decreases with increasing $\zeta$ (see~\Cref{tab:correlation_table}). 
    %(0.54) and the correlation becomes poor on further increasing the $\zeta$ values
    Similarly, the correlation of $m^*$ with $\bar{\Lambda}$ decrease with increasing $\zeta$ for both ${1.4M_{\odot}}$ (from 0.69 to 0.14) and for massive $2M_{\odot}$ NS (from 0.86 to 0.37). 
    \item We also notice a moderate correlation between $R_{1.4M_{\odot}}$ and $n_0$ increasing $\zeta$ from 0 to 0.03.
  %  \item At $\zeta=0.03$, $n_0$ shows moderate correlations with NS properties of a massive $2M_{\odot}$.
\end{enumerate}

\vskip 0.5 cm

\begin{figure*}[htbp]
    \centering
    \includegraphics[width=\linewidth]{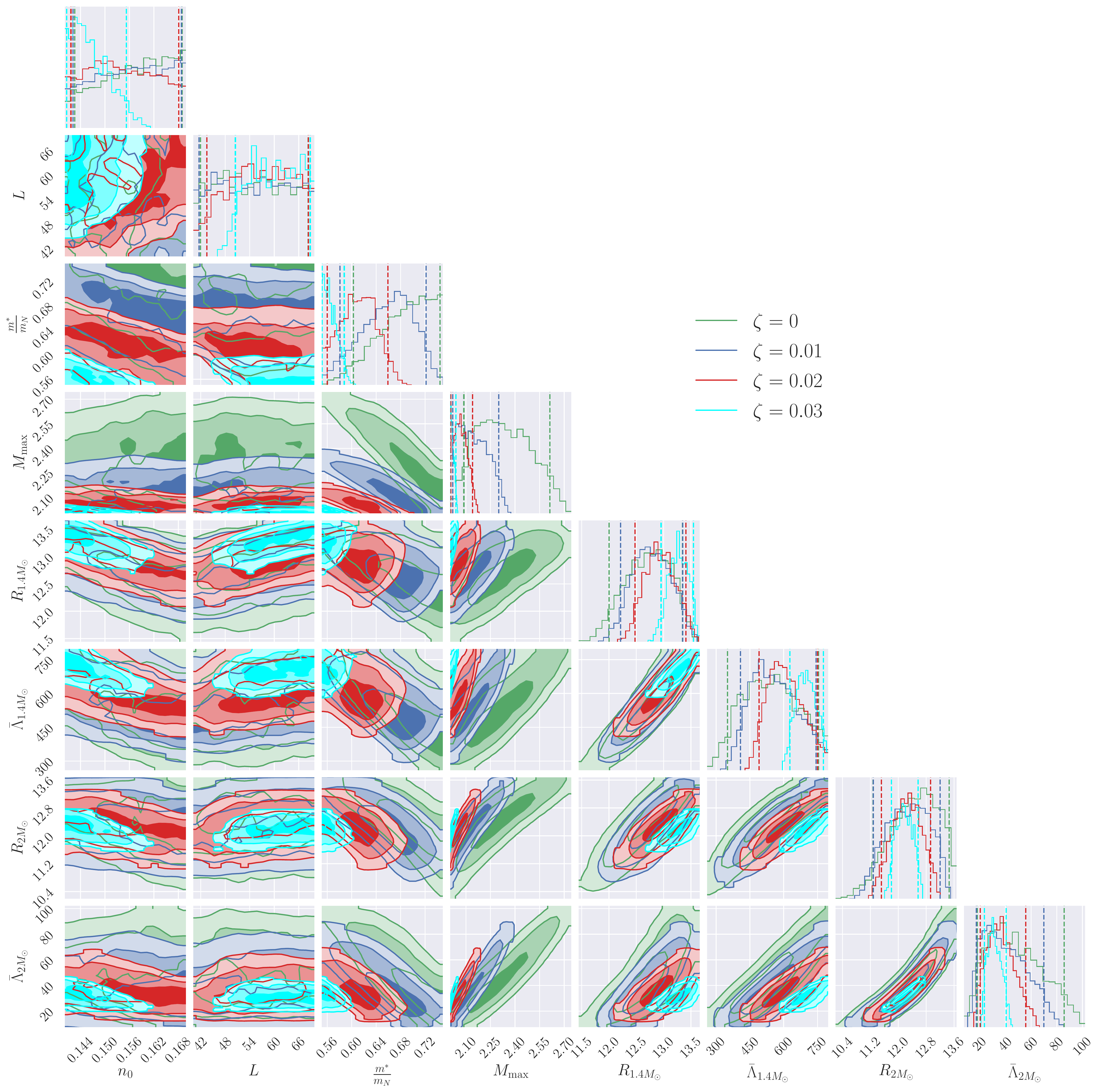}
    \caption{Joint posterior  distribution of saturation parameters and NS properties of a canonical $1.4M_{\odot}$ NS and a massive $2M_{\odot}$ NS. Different distributions corresponding to different constant $\zeta$ EoS models are shown with different colors.}
    \label{fig:distribution_at diff_zeta}
\end{figure*}

 We also check how the obtained posteriors for nuclear and NS parameters are affected by changing self vector interaction strength. The joint posterior distributions of nuclear parameters and NS properties for $\zeta \in [0,0.03]$ are displayed in~\Cref{fig:distribution_at diff_zeta}, from which the impact of vector self-interaction strength can be concluded as follows:
\begin{enumerate}
    \item In the absence of $\zeta$, higher $m^*$ values are more favourable (see the distribution of $m^*$ for $\zeta=0$ in ~\Cref{fig:distribution_at diff_zeta}). This is expected, as at $\zeta=0$ the $\bar{\Lambda}_{1.4M_{\odot}}$ constraint from GW170817 plays an important role ($2M_{\odot}$ constraint is mostly satisfied), favoring soft EoSs and higher effective masses corresponding to lower tidal deformability values. For $\zeta=0.01,0.02$, the EoSs being already softened due to the appearance of $\zeta$, higher $m^*$ (softer EoSs) become incompatible with the $2M_{\odot}$ constraints. Now the combined effect of GW170817 and maximum NS observed mass make the distribution of $m^*$ to peak at $0.66m_N$ and $0.61m_N$ for $\zeta=0.01$ and 0.02, respectively. Further increasing $\zeta$ to 0.03, the EoS models with $m^*>0.6m_N$ fail to produce a $2M_{\odot}$ NS and the distribution of $m^*$ shifts towards the lower $m^*$ values, i.e, $0.56m_N\leq m^* \leq 0.6m_N$.
    %\st{As the appearance of $\zeta$ further softens the EoS, the maximum mass constraint of massive pulsar will favour a lower value of $m^*$ (in constant $\zeta$ models, the stiffness is still controlled by $m^*$ ). For $\zeta=0.01,\ 0.02$ bound on the $m^*/m_N$ are found to be $0.66^{+0.04}_{-0.05}$ and $0.61^{+0.03}_{-0.03}$ respectively, which indicates a $\sim$8\% change in the median values. For $\zeta=0.03$, though we did not notice a peak in the $m^*$, we noticed that the lower $m^*/m_N$ are more favorable (as they produce stiff EoSs in the absence of $\zeta$) and found a 90\% upper bound at $m^*=0.57$. This implies that the favoured values of  $m^*$ for astrophysical constraints shifts from a higher value ($>$0.7) to a lower value ($\sim 0.57$) with increasing strength of vector self-interaction.}
    \item %\st{With increasing $\zeta$ from 0 to 0.01, the median of maximum mass that the model can support  changes from 2.32 to 2.14 which decreased by $\sim 7.75\%$. Further for $\zeta=0.02$, median of $M_{\rm max}$ shifts to $2.07$ (a decrease of ~10.77\% compared to $\zeta=0$ models).  Further increasing $\zeta$ to 0.03, we found a 90\% upper bound on the $M_{\rm max}\sim 2.03$.}
    The maximum stable NS mass ($M_{\rm max}$) that can be explained by EoS models with $\zeta=0$ is $\sim2.70M_{\odot}$, whereas, with finite $\zeta$ models with constant values of $\zeta=0.01,0.02$ and 0.03, the RMF models can support stable NSs with a maximum mass $M_{\rm max}$ $\sim 2.4M_{\odot},2.15M_{\odot}$ and $2.05 M_{\odot}$ respectively. In ~\Cref{fig:distribution_at diff_zeta}, the distribution of $M_{\rm max}$ for $\zeta=0.03$ is too narrow to be visible, this can be understood using the $M-R$ plane displayed in ~\Cref{subfig:mr_range_diffzeta} for $\zeta=0.03$.  
    \item At higher $\zeta$ values, where lower $m^*$ values become more probable, the allowed range (mainly the lower value) of the slope of symmetry energy ($L$) shifts to higher value and the lower $L$ values become less favored. %\st{(e.g, for $\zeta$=0.02, the 90\% lower bound of $L\sim 48 MeV $ and for $\zeta=0.03$ the lower bound shifts to $54$ MeV)}
    This generally happens due to the unphysical behavior of EoSs (unstable regions due to the appearance of negative pressures) at lower $m^*$ and lower $L$ values, and any such parameter set showing unphysical behavior is not considered for further analysis (also the window function present in Eq. ~\eqref{eqn:weight} becomes zero). This finding is consistent with the results reported in ~\cite{Hornick}. For $\zeta=0.03$ we find a strict bound on the lower limit of $L\sim 48$ MeV, i.e., $L$ should be $\geq48$ MeV. We also notice an  upper bound on $n_0\sim$ 0.160 fm$^{-3}$ at $\zeta=0.03$. The increasing correlation at a higher $\zeta$ value at 0.03 can be understood from Fig.\ref{fig:eos_mr_range_diff_zeta}:  the spread in the EoS appears at a lower density regime (1-2.5 $n_0$) while at high density the maximum mass constraint narrows down the EoS as well as the mass-radius range. With increasing $\zeta$, the width of the sample distribution of NS observables becomes more constrained.
    \item %\st{The median, 90\% credible interval for $R_{1.4M_{\odot}}$  for $\zeta=0,\ 0.01,\ 0.02$ and 0.03 are found to be $12.74^{+0.41}_{-0.48}$, $12.77^{+0.39}_{-0.36}$, $12.93^{+0.31}_{-0.29}$ and $13.27^{+0.19}_{-0.19}$ respectively.} 
    The distribution of NS properties for both canonical $1.4M_{\odot}$ and massive $2M_{\odot}$ becomes more constrained with increasing $\zeta$ values ( see~\Cref{fig:distribution_at diff_zeta} and ~\Cref{subfig:mr_range_diffzeta} also). The peak of distribution of $R_{1.4M_{\odot}}$ and $\bar{\Lambda}_{1.4M_{\odot}}$ shifts towards higher values with increasing $\zeta$, while, for  massive $2M_{\odot}$ NSs the peak of distribution of $R_{2M_{\odot}}$ and $\bar{\Lambda}_{2M_{\odot}}$ moves  towards lower values.
    
\end{enumerate}

\begin{table*}
\caption{Variation of correlation coefficients with different fixed $\zeta$ models.}
\label{tab:correlation_table}
\begin{tabular}{| c | c | c | c | c | c | c | c | }
%\hline & & & & & & &  \\
\hline
$\zeta$ & $L$-$R_{1.4M_{\odot}}$ & $L$-$R_{2M_{\odot}}$ & $m^*$-$M_{max}$ & $m^*$-$R_{1.4M_{\odot}}$ & $m^*$-$\bar{\Lambda}_{1.4M_{\odot}}$ & $m^*$-$R_{2M_{\odot}}$ & $m^*$-$\bar{\Lambda}_{2M_{\odot}}$\\
\hline
    0.00 & 0.3 & 0.04 & 0.93 & 0.54 & 0.69 & 0.78 & 0.86\\
\hline
    0.01 & 0.42 & 0.15 & 0.91 & 0.43 & 0.60 & 0.77 & 0.84\\
\hline
    0.02 & 0.44 & 0.19 & 0.78 & 0.13 & 0.30 & 0.59 & 0.67\\
\hline
    0.03 & 0.60 & 0.08 & 0.49 & 0.19 & 0.14 & 0.27 & 0.37\\
\hline
\end{tabular}

\end{table*}

\subsection{Maximum neutron star mass and $\omega$-meson self interaction strength ($\zeta$)}\label{subsec:limit_zeta}

\Cref{fig:eos_mr} shows that the introduction of $\omega$-meson self-interaction efficiently softens the high-density behaviour of the EoS. The maximum NS mass that can be supported by an EoS primarily depends on the stiffness (or softness) of the EoS, which is affected by the high-density property of matter. Any correct theoretical EoS model should be able to explain the highest observed neutron star mass ($\geq 2M_{\odot}$). 
%This change controls high-density behaviour by $\zeta$, which is reflected in the stable maximum NS mass~(\Cref{subfig:MR_diff_zeta}). 
By applying the maximum mass constraint resulting from the observed massive pulsar, one can limit the allowed range for $\zeta$. Other physical constraints considered in different works include the constraints following chiral effective field theory ($\chi$EFT), which describes the behavior of low-density nuclear matter (0.5-1.4$n_0$)~\cite{Most,Malik2017,Essick2020,Ghosh2022,Ghosh2022b,Patra2022,Alford2022}. At such low densities, the nuclear matter is insensitive to $\zeta$ ~\cite{MULLER1996}. Moreover, we note that the behavior of meson fields in the regime of (0.5$n_0$-1.4$n_0$) is relatively independent of $\zeta$ ~(\Cref{fig:meson_fields}). This scenario allows $\zeta$ to be constrained solely by applying the maximum observed mass constraint.
\\

\begin{figure*}[htbp]
\centering
\begin{subfigure}{.49\textwidth}
  \centering
  \includegraphics[width=\linewidth]{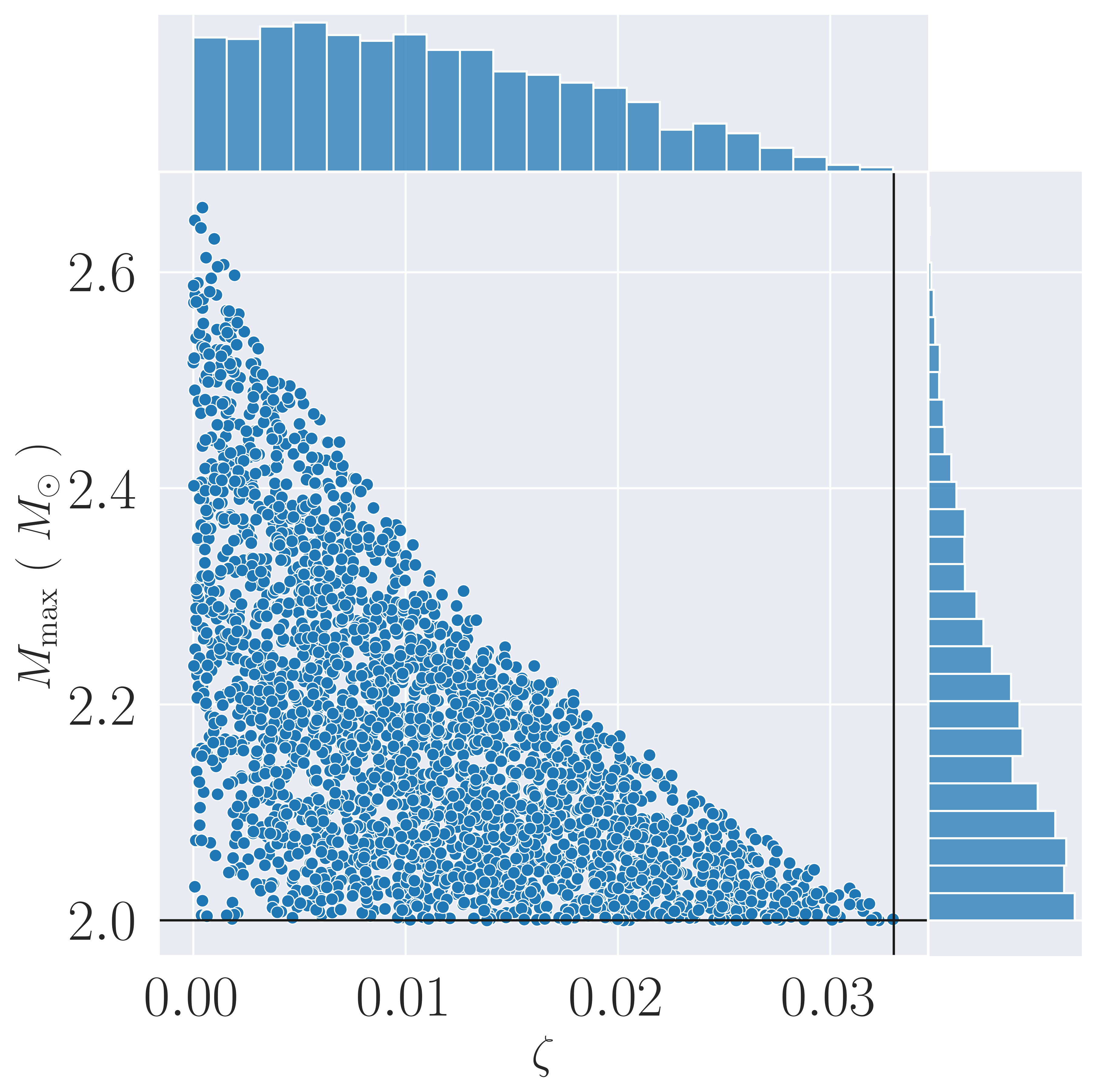}
  \caption{}  
  \label{subfig:zeta_mmax}
\end{subfigure}%
\begin{subfigure}{.49\textwidth}
  \centering
  \includegraphics[width=\linewidth]{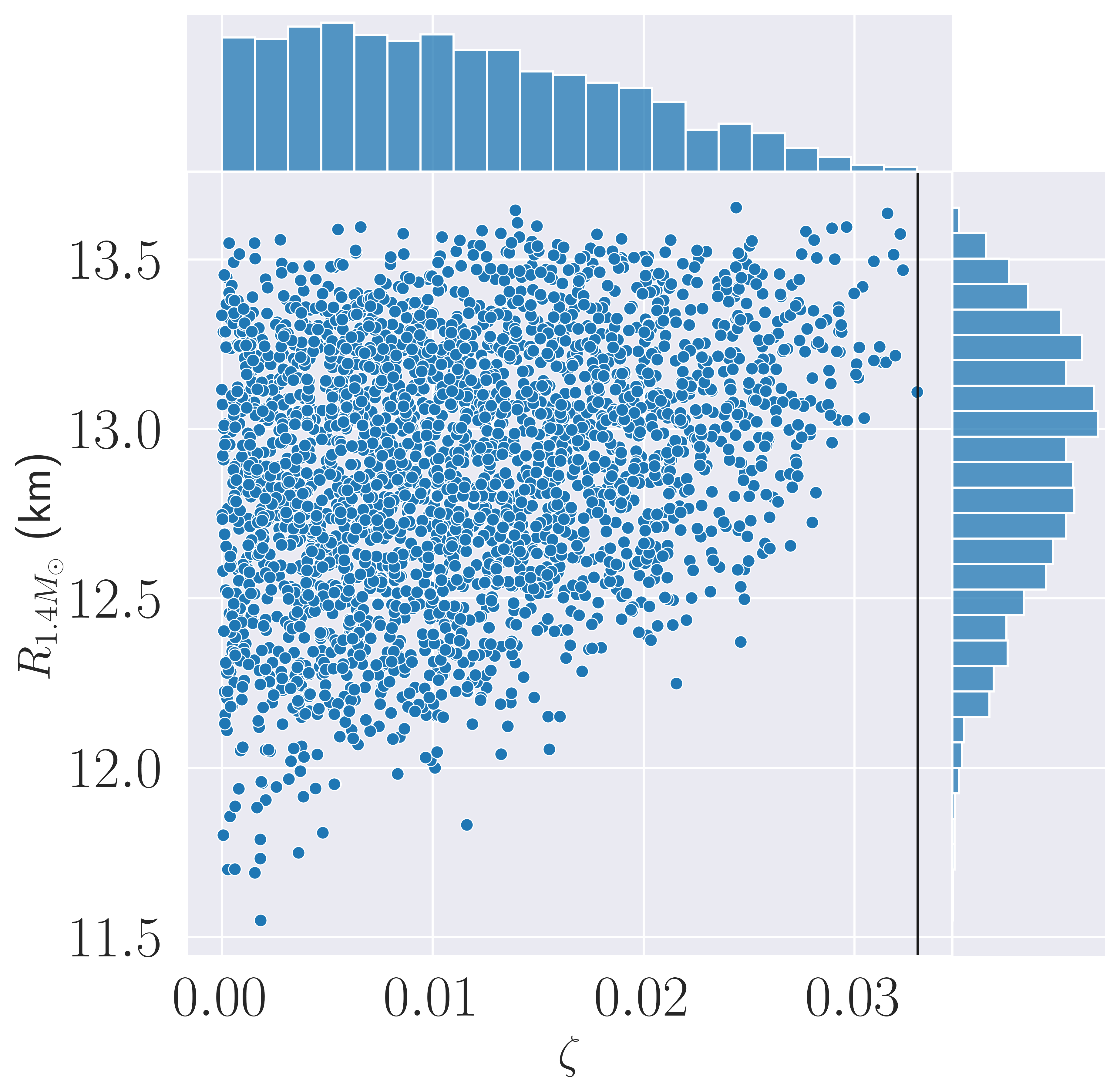}
  \caption{}
  \label{subfig:zeta_r1.4}
\end{subfigure}
\caption{(a) Distribution of $\zeta$ and the maximum mass ($M_{\rm max}$). The horizontal black line corresponds to the maximum $2M_{\odot}$ constraint, while the vertical black line corresponds to the maximum limiting value of $\zeta$ after applying  the $2M_{\odot}$ constraint. (b) Distribution of $\zeta$ and the radius of a canonical NS  ($R_{1.4M_{\odot}}$).}
\end{figure*}

\begin{figure}[htbp]
    \centering
    \includegraphics[width=\linewidth]{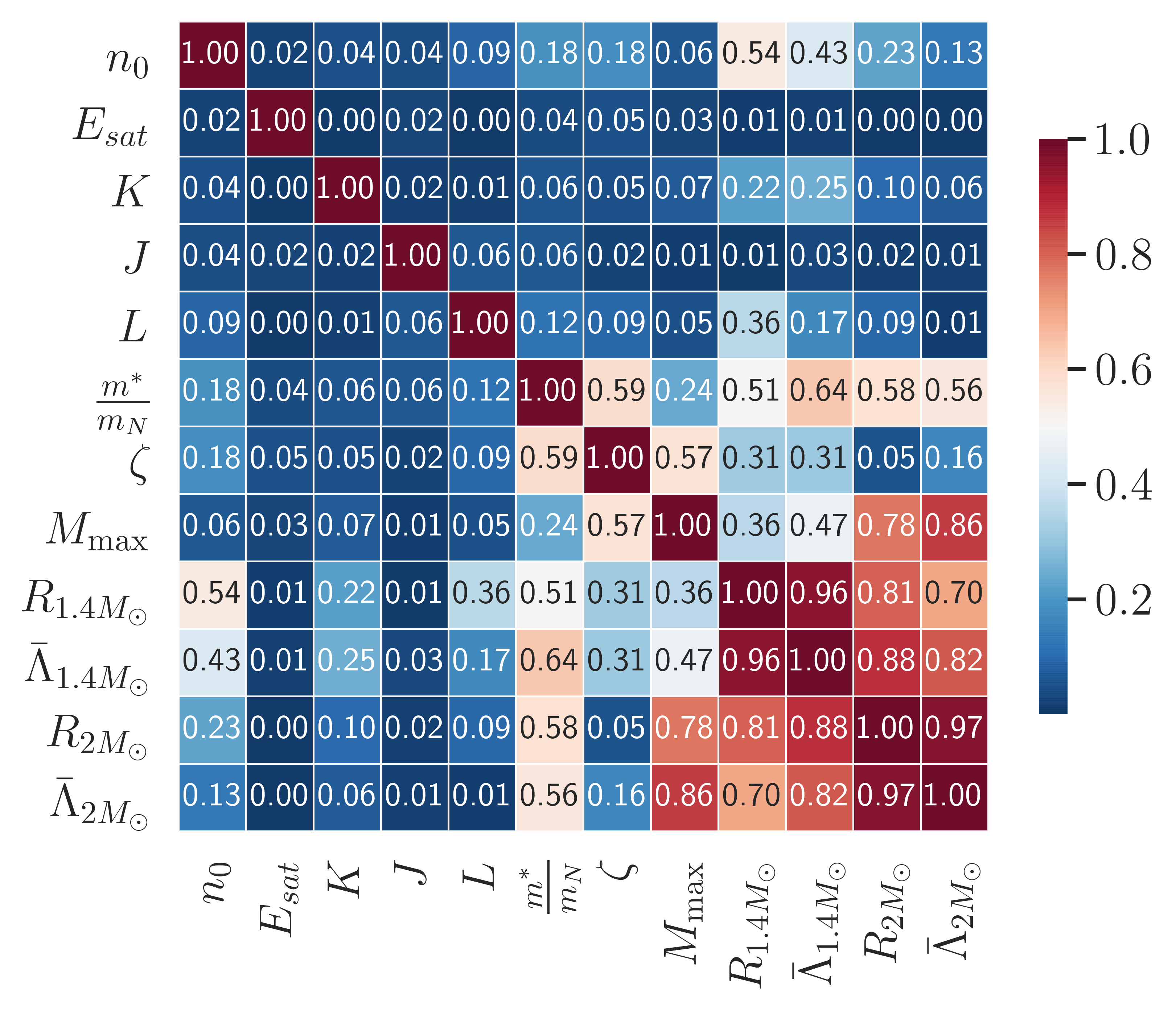}
    \caption{Correlation matrix showing correlations among nuclear saturation parameters themselves as well as with NS properties. The  saturation parameters are varied in the range given in `variation' row of ~\Cref{tab:HTZCS_parameter_set}.  }
    \label{fig:correlation_zeta_uniform}
\end{figure}

\begin{figure*}[htbp]
    \centering
    \includegraphics[width=\linewidth]{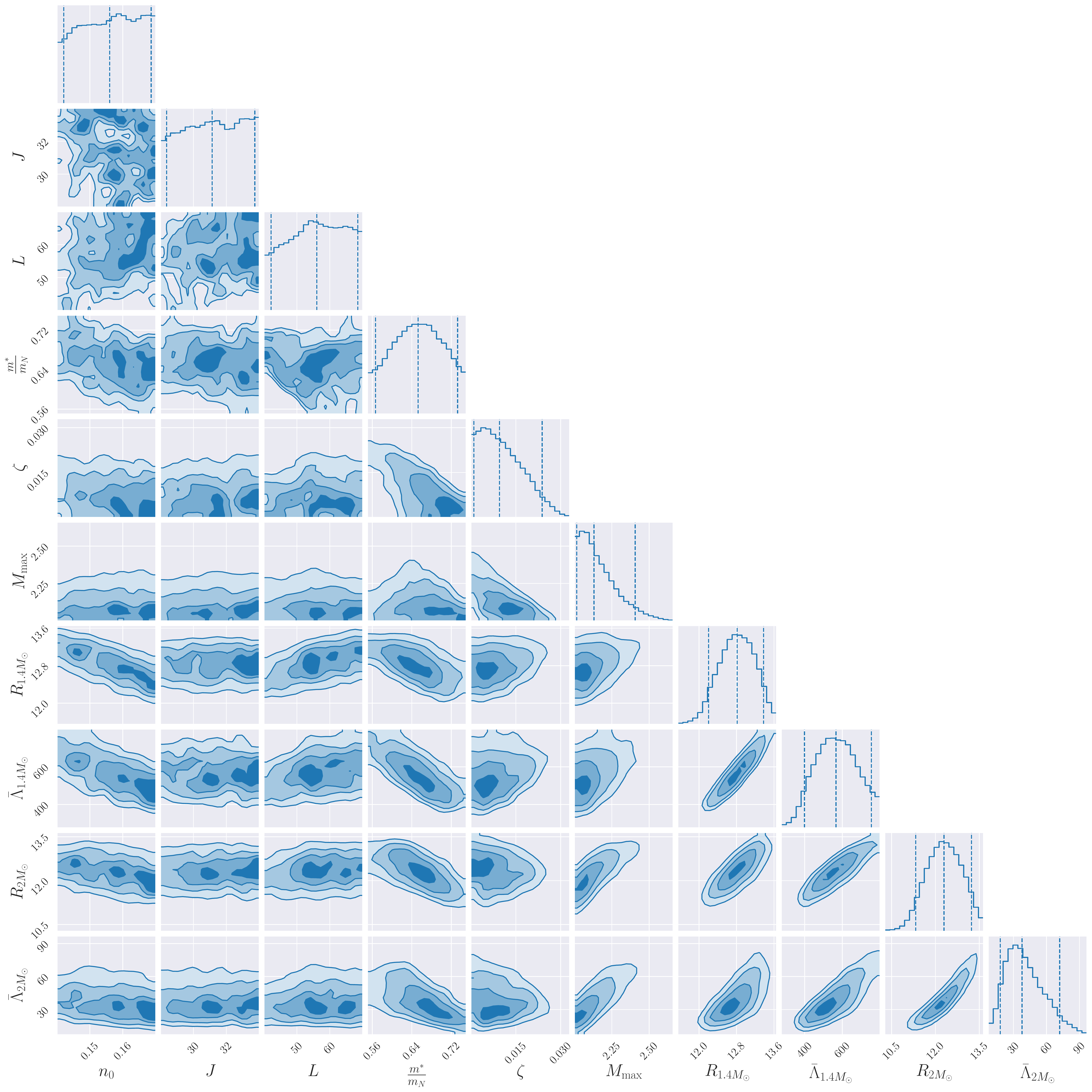}
    \caption{Joint posterior  distribution of saturation parameters and NS properties of a canonical $1.4M_{\odot}$ NS and a massive $2M_{\odot}$ NS. Prior for $\zeta$ is set to be a uniform distribution.}
    \label{fig:distribution_diff_zeta}
\end{figure*}

 The model parameters in the framework of the RMF model are fixed to nucleon saturation parameters and subject to measurement uncertainties. In addition to considering the random uniform prior distribution for the nuclear saturation parameters in the threshold range of the range parameters~\Cref{tab:HTZCS_parameter_set}, we now assume a uniform prior for $\zeta \in [0,0.1]$. Based on the maximum observed mass constraints, we are able to put a maximum limit on $\zeta$ close to 0.033 (i.e., $\zeta \leq 0.033$. Also see ~\Cref{subfig:zeta_mmax}). Additionally, we display the distribution of $R_{1.4M_{\odot}}$ in~\Cref{subfig:zeta_r1.4}. Our EoSs are compatible with the upper limit of tidal deformability obtained from the analysis of GW170817~\cite{Abbott2019}. Based on previous studies, it has been shown that the nucleon effective mass controls the stiffness of EoS in the absence of $\zeta$ and has shown a notable correlation with the NS maximum mass. After including the self vector interaction, the favorable value for nucleonic effective mass and correlation will be altered. A discussion on how a fixed vector self-interaction strength impacts other saturation parameters and their correlation with NS properties is presented in~\Cref{sec:correlation}. Using the statistical weighting factor for EoS parameters corresponding to the maximum observed pulsar mass PSR J0740+6620 and the tidal deformability constraint from the BNS event GW170817, the correlation matrix is displayed in ~\Cref{fig:correlation_zeta_uniform} and the posterior distributions for some saturation parameters and the NS properties of a canonical 1.4$M_{\odot}$ and $2M_{\odot}$ are presented in~\Cref{fig:distribution_diff_zeta}. Comparing the correlation matrix~\ref{fig:correlation_zeta_uniform} and posterior distributions~\Cref{fig:distribution_diff_zeta} with the correlation matrix (~\ref{subfig:corr_zeta_0})  and posterior distribution (~\ref{fig:distribution_at diff_zeta})  corresponding to EoS models with $\zeta=0$, one can conclude the following:
 
 \begin{itemize}
     \item The strong correlation of $m^*$ with $M_{\rm max}$ at $\zeta=0$ becomes poor and drops to 0.24 after considering the variation of $\zeta$. This correlation is much lower compared to the minimum correlation among $m^*$ and $M_{\rm max}$ (0.49) for $\zeta=0.03$ constant EoS models. The decreasing  correlation due to the appearance of $\zeta$ can be understood as $\zeta$ also plays a vital role in controlling the high-density behavior of NS EoS.
     \item $\zeta$ shows a moderate correlation with $m^*$ (0.59) as well as with $M_{\rm max}$ (0.57). We do not notice any significant correlation of $\zeta$ with other saturation parameters or with any other NS properties.
     \item Comparing with $\zeta=0$ EoS models, the correlation of $m^*$ with properties of a $1.4M_{\odot}$ NS do not change significantly. However, the strong  correlation of $m^*$ with NS properties of a $2M_{\odot}$ at $\zeta=0$, drops to moderate  (0.58 for $R_{2M_{\odot}}$ and 0.56 for $\bar{\Lambda}_{2M_{\odot}}$)  by allowing the variation of $\zeta \in [0,0.1]$.
     \item $L$ shows a poor correlation with $R_{1.4M_{\odot}}$ (0.36). We notice a moderate correlation of 0.54 among  $n_0$ and  $R_{1.4M_{\odot}}$.
     \item Looking at the distribution of $m^*$ from   ~\Cref{fig:distribution_diff_zeta} with $\zeta \in [0,0.1]$ the favoured value of $m^*$ peaks $\sim 0.65m_N$ compared to the $m^*$ distribution in ~\Cref{fig:distribution_at diff_zeta} for $\zeta=0$ where higher $m^*$ values are more favourable. From the distribution of $\zeta$ values, it is evident that lower $\zeta$ values are more favored as they can produce a large maximum mass above the observed maximum mass constraints.
     %\st{the median and the 90\% credible region for $m^*$ is found to be $0.65^{+0.06}_{-0.06}$. The favored value of  $m^*$ shifts   0.65 compared to the $\zeta=0$ case, where the higher $\zeta$ values are highly favorable. Though we did not notice a peak in the distribution of $\zeta$ values, the lower $\zeta$ values are more favorable as they can produce a large maximum mass above the observed maximum mass constraints.}
     \item Imposing the astrophysical constraints, we found the upper limit on $\zeta$ to be $\sim 0.033$.
 \end{itemize}
 \begin{center}

\end{center}

%%%%%%%%%%%%%%%%%%%%%%%%%%%%%%%%%%%%%%%%
\section{Implications}
\label{sec:discussions}

In this work, we investigated the importance of vector self-interaction term $\zeta$ within the RMF model framework in determining the NS observable properties such as mass, radius, and tidal deformability. For this, we first investigated the effect of this non-linear term on the global properties of a fixed nuclear parametrization. We verified that the maximum NS mass decreases with increasing $\zeta$. This is in accordance with the understanding that the $\omega$-meson self-interaction results in softening of the EoS at high-density \cite{Horowitz2001,MULLER1996}, which explains the fact why this term was not included in many recent studies employing the RMF model in order to maintain compatibility with a high maximum mass of 2 M$_{\odot}$. For different parametrizations, variation of the effective nucleon mass $m^*/m_N$ and the vector self-interaction $\zeta$ permit a modification of the high-density component of the EoS. We further investigated the behavior of the meson fields as a function of density for different $\zeta$. We found that the vector $\omega$ meson field behavior, changes from $\omega \propto n$ for $\zeta =0$ to $\omega \propto n^{1/3}$ at large densities for $\zeta \neq 0$, which is in accordance with the findings of \cite{Horowitz2001,MULLER1996}. A similar behavior is seen in the $\rho$ field (symmetry energy) in pure neutron matter. \\

We then performed a Bayesian analysis, by initially varying the nuclear empirical parameters within their present uncertainties and investigated possible correlations among the parameters and NS observables for different constant values of $\zeta$.  With increasing $\zeta$, the width of the sample distribution of NS observables becomes more constrained. We found that the correlation between the slope of the symmetry energy $L$ and the radius $R_{1.4 M_{\odot}}$ increases as $\zeta$ goes from 0 to higher non-zero values, and we notice a maximum correlation of 0.6 for $\zeta=0.03$. These results can be compared with Hornick et.~al. \cite{Hornick} where $\zeta=0$ and there is negligible dependence of the slope parameter on the radius, as the $\rho$ field decreases with the density as $\rho \propto 1/n$ for $\zeta =0$. The correlation of $m^*$ with NS observables also decreases with increasing $\zeta$. For constant $\zeta$ models, we found that the favored range for $m^*$ shifts from higher to lower values on increasing $\zeta$ from 0 to 0.03. At higher $\zeta$ values, as lower $m^*$ becomes more probable, lower $L$ values become less favored. For $\zeta=0.03$ constant RMF model, we find a lower bound on $L$ nearly 48 MeV and an upper bound on $n_0\sim 0.160 \rm \  fm^{-3}$.
\\
%\st{ We notice an increase of 4\% ( 28.75\%) in the median value of $R_{1.4M_{\odot}}$ ($\bar{\Lambda}_{1.4M_{\odot}}$) by changing $\zeta$ from 0 to 0.03. However, for a massive $2{{M_{\odot}}}$  NS with models at $\zeta=0.03$ we found a decline of $\sim 3$\% (32\%) in the median of $R_{2M_{\odot}}$ ($\bar{\Lambda}_{2M_{\odot}}$) compared to models with $\zeta=0$.}

In order to set a maximum possible value of the vector self-interaction strength, we vary $\zeta$ within a uniform prior [0,0.1] in the Bayesian analysis, along other saturation parameters within the range given in ~\Cref{tab:HTZCS_parameter_set}. On demanding compatibility with astrophysical observations, we found that the maximum value of the vector self-interaction cannot exceed 0.033. This therefore also sets the maximum dependence of the correlation of $L$ with $R_{1.4 M_{\odot}}$ ($\sim 0.60$) at $\zeta=0.03$. The maximum restricted value of $\zeta \sim0.033$ explains our consideration of $\zeta=0.03$ as the maximum value in the systematic investigation explained in ~\Cref{subsec:corr_constant_zeta}.
For the statistically weighted study including the variation of $\zeta$, the correlation of $m^*$ with maximum mass $M_{\rm max}$ drops to a low value in contrast, while for constant $\zeta$ RMF models we observed a decreasing correlation between $m^*$ and $M_{\max}$ with increasing $\zeta$. The vector self-interaction strength shows a moderate correlation with $m^*$ and $M_{\rm max}$, and no significant correlations with other nuclear or NS parameters. Interestingly, for models with varying $\zeta$, $m^*$ only shows moderate correlations with NS properties compared with strong correlations with NS properties at fixed $\zeta$ models.  The correlation among symmetry energy $L$ and $R_{1.4M_{\odot}}$ do not change significantly and remains $\sim$ 0.36 after considering the variation of vector self-interaction strength compared  to models ignoring the vector self-interaction (though fixing the vector self-interaction strength near its maximum value at $\zeta=0.03$ predicts a high correlation of 0.6 among $L$ and $R_{1.4M_{\odot}}$). \\

%\st{For the statistical weighted study including the variation of $\zeta$, the strong correlation of $m^*$ with maximum mass drops to a low value. The vector self-interaction strength shows a moderate correlation with $m^*$ and $M_{\rm max}$, and no significant correlations with other nuclear or NS parameters. Interestingly, for models with varying $\zeta$, the $m^*$ only shows moderate correlations with NS properties compared to strong correlations with NS properties at fixed $\zeta$ models. From the posteriors obtained from the Bayesian study, we imposed the astrophysical constraints of maximum mass and tidal deformability to restrict the range of the $\zeta$ parameter: we found that in order to be compatible with observations, the maximum value of the vector self-interaction cannot exceed 0.033. This therefore also sets the maximum dependence of the correlation of $L$ with $R_{1.4 M_{\odot}}$ ($\sim 0.60$).}  \\

The results of this investigation are relevant and timely, as currently, many works are probing possible correlations between nuclear empirical parameters and NS multi-messenger properties, such as that between the slope of the symmetry energy and the radius of a canonical NS. This study shows that vector self-interaction plays an important role not only in controlling the high-density properties of the nuclear EoS, but also in governing such relations. This explains why a strong correlation is found in certain cases, while a weak correlation is found in others: the correlations depend on the applied ansatz for vector self-interaction. The maximum strength of the vector self-interaction obtained from state-of-the-art NS observational data also sets an upper limit to the strength of such a correlation.

\section*{Acknowledgements}
B.-K.P. and D.C. acknowledge the usage of the IUCAA HPC computing facility for numerical calculations. J.S.-B. acknowledges support by the Deutsche Forschungsgemeinschaft (DFG, German Research Foundation) through the CRC-TR 211 `Strong-interaction matter under extreme conditions’—project number 315477589—TRR 211. B.K.P thanks  Suprovo Ghosh, Swarnim Shirke and Dhruv Pathak for the useful discussions they have during this work. B.K.P also thanks Liam Brodie for the useful discussions they have had regarding this work. The authors thank the anonymous reviewer for the insightful suggestions made during the review process to improve the article.
% Uncomment and use as the case may be
%\begin{theorem} 
%\end{theorem}

% Uncomment and use as the case may be
%\begin{lemma} 
%\end{lemma}

\appendix

\section{Imposing $\chi \rm EFT$ constraint}\label{sec:chieft_withzeta}

\begin{figure*}[htbp]
\centering
\begin{subfigure}{.49\textwidth}
  \centering
  \includegraphics[width=\linewidth]{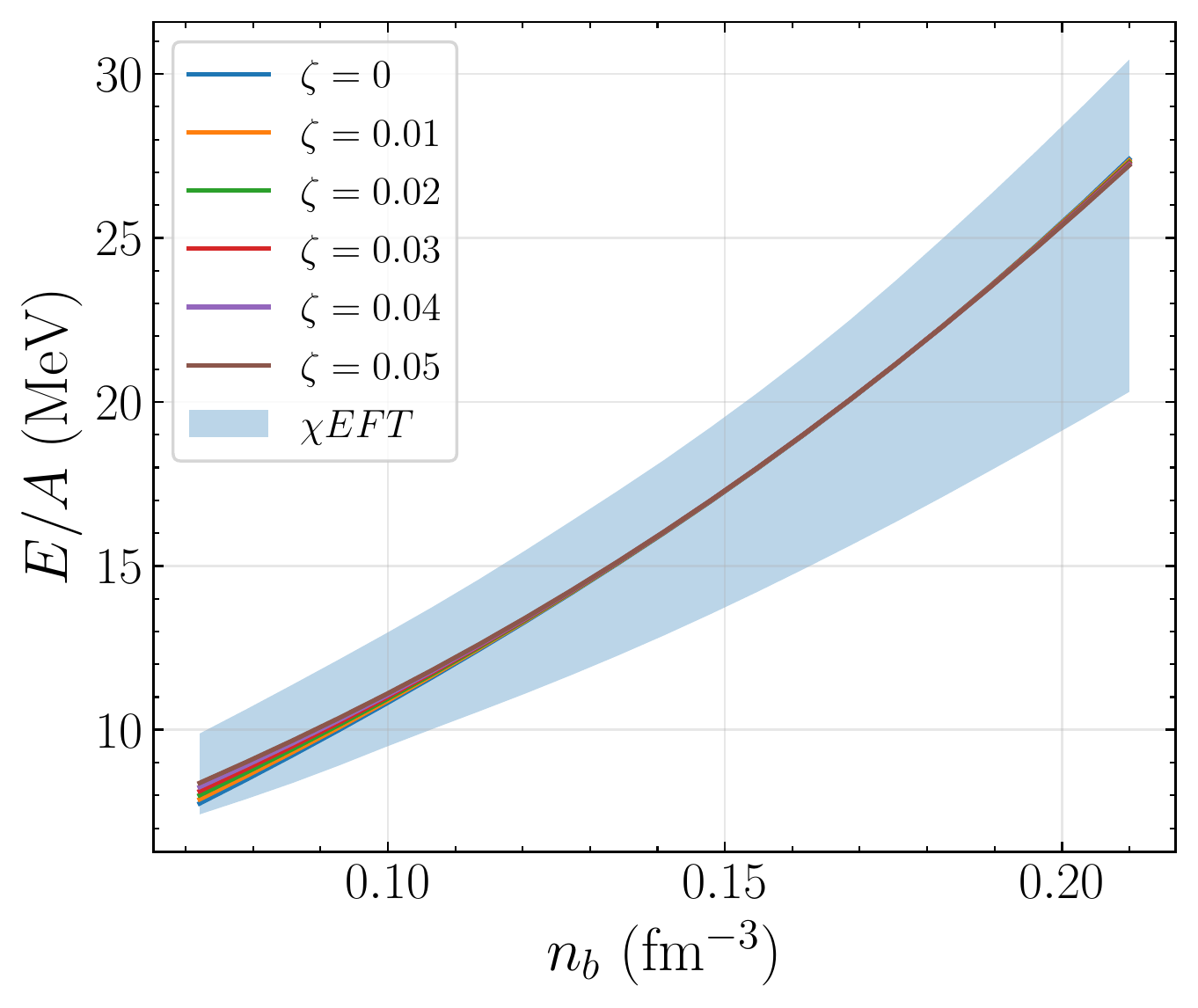}
  \caption{}  
   \label{fig:PNM_diff_zeta}
\end{subfigure}%
\begin{subfigure}{.49\textwidth}
  \centering
  \includegraphics[width=\linewidth]{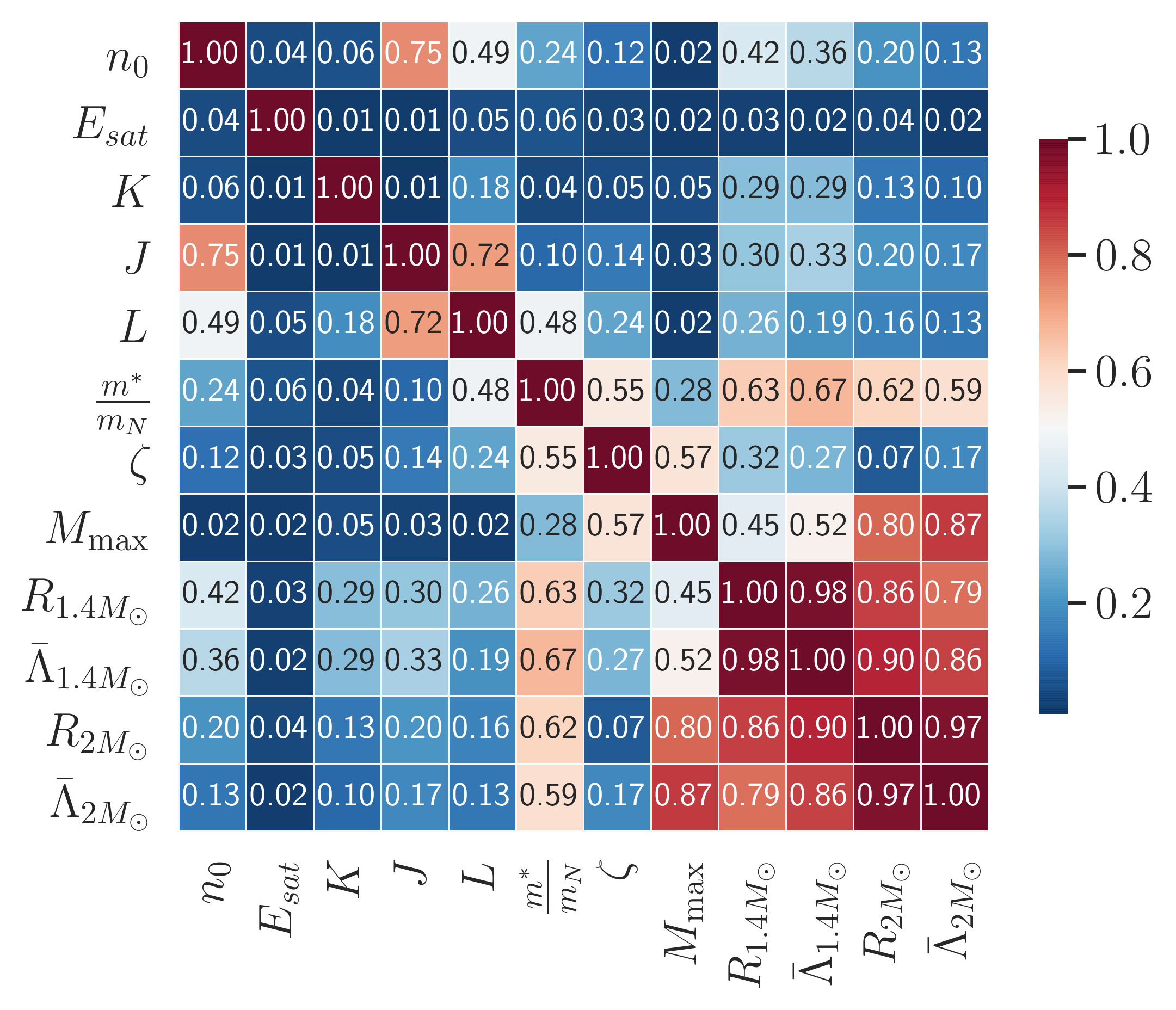}
  \caption{}
  \label{fig:corr_astrochieft}
\end{subfigure}
\caption{(a) Binding energy E/A of selective EoS model from ~\Cref{tab:HTZCS_parameter_set} for pure neutron matter (PNM)
as a function of baryon  density $n_b$ along with the 
chiral effective field theory ($\chi \rm EFT$) data
~\cite{Drischler}. (b)Correlation matrix after considering $\chi \rm EFT$ constraints along with astrophysical constraints. The RMF model includes the vector self interaction and the uncertainty range for parameters are set from ~\Cref{tab:HTZCS_parameter_set}. }
   
\end{figure*}

As discussed in the ~\Cref{sec:prelim}, $\zeta$ does not significantly impact the nuclear matter properties in the density regime $0.5n_0\leq n_b\leq1.5n_0$ where  $\chi \rm EFT$ plays a crucial role constraining the nuclear parameters. In Ghosh et al.~\cite{Ghosh2022}, it is shown that after imposing the $\chi \rm EFT$ constraint along with the astrophysical conditions for RMF models without considering the vector self-interaction ($\zeta=0$), some correlations among the nuclear parameters change significantly. The detailed analysis with the implication of astrophysical and $\chi \rm EFT$ constraints for $\zeta=0$ RMF model is done in ~\cite{Ghosh2022}. Here we will discuss the impact of $\chi \rm EFT$ constraint considering the RMF model with allowing the variation of $\zeta \in [0,0.1]$ along with the variation of other saturation parameters from ~\Cref{tab:HTZCS_parameter_set}. We use the data of binding energy of the PNM matter reported in ~\cite{Drischler} to calculate the weight for a parameter set $\{P\}$ as discussed in ~\Cref{sec:correlation}. The correlation matrix on applying the $\chi \rm EFT$ constraints on top of astrophysical constraints is displayed in~\Cref{fig:corr_astrochieft}. Comparing the correlation matrices ~\Cref{fig:correlation_zeta_uniform} and ~\Cref{fig:corr_astrochieft}, one can conclude that after imposing the $\chi \rm EFT$ constraint, the correlation among the symmetry energy ($J$) and slope of symmetry energy ($L$) becomes strong (0.72) compared to the negligible correlation of 0.02 due to imposition of the astrophysical constraint only. We also notice a strong correlation of 0.75 among $n_0$ and $J$ after imposing the $\chi \rm EFT$ constraint. These findings are consistent with \cite{Ghosh2022} (see Fig. 7 therein). We do not see any significant change in the correlations of nuclear saturation parameters with the NS observables after the implication of $\chi \rm EFT$ constraints.

\printcredits

\bibliographystyle{elsarticle-num}
%\bibliographystyle{cas-model2-names}

\bibliography{Pradhan}



\end{document}